\documentclass{aa}  
\usepackage{natbib}
\bibpunct{(}{)}{;}{a}{}{,}
\usepackage[normalem]{ulem}
\usepackage{xcolor}
\usepackage{siunitx}

\usepackage{graphicx}
\usepackage{txfonts}

\begin{document} 

   \title{
   Evolution and Mass Dependence of UV-to-near-IR Color Gradients up to $z=2.5$ from HST$+$JWST}
   \titlerunning{UVJ color gradients across $0.5<$~z$~<2.5$}

   \author{M. Martorano\inst{1}
          \and
          A. van der Wel\inst{1}
          \and
          A. Gebek \inst{1}
          \and
          M. Baes \inst{1}
          \and
          E. F. Bell \inst{2}
          \and
          G. Brammer\inst{3}
          \and
          S. E. Meidt\inst{1}
          \and
          A. Nersesian\inst{4,1}
          \and
          K. Whitaker\inst{3,5}
          \and
          S. Wuyts \inst{6}
          }

   \institute{
             Sterrenkundig Observatorium, Universiteit Gent, Krijgslaan 281 S9, 9000 Gent, Belgium\\
             \email{marco.martorano@ugent.be}
        \and
             Department of Astronomy, University of Michigan, 1085 South University Avenue, Ann Arbor, MI 48109–1107, USA
        \and
             Cosmic Dawn Center (DAWN), Niels Bohr Institute, University of Copenhagen, Jagtvej 128, København N, DK-2200, Denmark
        \and
             STAR Institute, Université de Liège, Quartier Agora, Allée du six Aout 19c, B-4000 Liege, Belgium
        \and
             Department of Astronomy, University of Massachusetts, Amherst, MA 01003, USA
        \and
            Department of Physics, University of Bath, Claverton Down, Bath BA2 7AY, UK
             }

   \date{Received 16 June, 2025; accepted 28 November, 2025}

  \abstract
   {}
   {We present the redshift evolution of radial color gradients (in rest-frame ${\textit{U}} - {\textit{V}}$ and ${\textit{V}} - {\textit{J}}$) for galaxies in the range $0.5<$~z~$<2.5$ and investigate their origin and dependence on stellar mass.}
   {We select $\sim 10,200$ galaxies with stellar masses $M_\star>10^{9.5}~{\text{M}}_\odot$ from publicly available JWST/NIRCam-selected catalogs. Using 2D S\'ersic profile fits to account for PSF broadening, we perform spatially resolved SED fitting on HST and JWST/NIRCam photometry retrieving accurate rest-frame ${\textit{U}} - {\textit{V}}$ and ${\textit{V}} - {\textit{J}}$ color gradients within 2$R_\text{e, F444W}$.}
   {Star-forming galaxies generally exhibit negative ${\textit{V}} - {\textit{J}}$ color gradients that are strongly mass and redshift dependent. For massive star-forming galaxies ($M_\star>10^{10.5}~{\text{M}}_\odot$) at $z>1.5$ ${\textit{V}} - {\textit{J}}$ colors are $\approx 0.5$~mag redder within the effective radius than outside, on average.
   We find that, at all redshifts and across the entire stellar mass range, ${\textit{V}} - {\textit{J}}$ gradients strongly correlate with global attenuation ($A_V$), suggesting that they predominantly trace dust attenuation gradients. Edge-on galaxies are redder and have stronger gradients at all $z$, although the correlation weakens at higher $z$.
   The ${\textit{U}} - {\textit{V}}$ and ${\textit{V}} - {\textit{J}}$ color gradients in the quiescent galaxy population, in contrast, are weakly negative (from $\approx -0.1$ to $\approx- 0.2$~mag), though significant, and show little or no dependence on stellar mass, redshift or axis ratio. The implication is that quiescent galaxies must be largely transparent, with low $A_V$, and color gradients mostly attributable to stellar population gradients.}
    {}

   \maketitle

\section{Introduction}\label{sec:Introduction}

Galaxies in the local universe show red centers and bluer outskirts \citep[i.e.,][]{lin17, ellison18} a characteristic attributed to gradients in the age and composition of stellar populations at different radii, as well as attenuation by dust. These color gradients have been deeply investigated since the early seventies \citep{sandage72} and their characteristics have been studied by many over the years \citep{peletier90, de-jong96a, guo11, szomoru13, wuyts13, liu16, liu17, suess19b, miller23, van-der-wel24}.
The color gradient in nearby quiescent galaxies can primarily be attributed to a radial metallicity gradient \citep[i.e.,][]{wu05,tortora10,tortora11,sanchez-blazquez14,goddard17,lin24} with the additional possibility of an age gradient \citep{la-barbera09}.

Color gradients in present-day star-forming galaxies arise from various factors \citep{bell00}. The inner parts have older and more metal-rich stellar populations \citep{bell00, zibetti22}, have lower star-formation activity \citep{tacchella16, lin17, belfiore17, belfiore18, ellison18, lin19}, and higher levels of attenuation \citep{greener20}. These factors are all physically correlated, but it is important to keep in mind that the origin of color gradients can vary from galaxy to galaxy: whereas the bulge of M31 has no ongoing star formation, other nearby spirals such as M51 and M101 have significant star formation throughout.

These results prompted an investigation of how these gradients emerged by looking at galaxies at earlier times, at higher redshift. The \textit{Hubble} Space Telescope (HST) led the way to unveiling the rest-frame optical colors of galaxies up to redshift $z\sim1$ \citep[see, e.g.,][]{hinkley01, menanteau01, mcgrath08}, showing color gradients were present also in those galaxies when the universe was $\sim6$~Gyr old. The limited rest-frame optical wavelength range of these initial observations creates difficulties for disentangling dust and stellar population gradients: galaxies (or galaxy regions) can appear redder because of old/metal-rich stellar populations or due to increased dust attenuation.
Near-IR emission at wavelengths between 1 and 3$~\mu{\text{m}}$, on the other hand, is less affected by dust attenuation and, therefore, allows us to separate between the effects of dust and stellar population gradients.
Several authors \citep[see e.g.,][]{wuyts12, szomoru13, liu16, wang17, liu17, suess20} have reported strong negative color gradients in high redshift galaxies, that is, their centers are redder than their outskirts. These gradients notably imply smaller half-light radii at longer wavelengths, and also more compact stellar mass distributions than stellar light distributions \citep[i.e.,][]{mcgrath08, dutton11, guo11, wuyts12, szomoru13, mosleh17, mosleh20, suess19a, suess19b, suess20, miller23, van-der-wel24}.
However, the impact of dust on these trends has been difficult to confirm without a near-IR view, which is not obtainable with HST above redshift $z>0.4$. 

The \textit{James Webb} Space Telescope NIRCam instrument solved this issue: its high angular resolution at near-IR wavelengths, comparable to that of HST in the optical, now allows us to investigate the rest-frame $>1~\mu\text{m}$ wavelength range for galaxies up to redshift $\sim2.5$, when the universe was just $\approx2.5~\text{Gyr}$ old. Early JWST results on a small sample of galaxies (54 star-forming galaxies with $M_\star>10^{10}~\text{M}_\odot$ and $1.7<z<2.3$) showed that galaxies exhibit strong negative color gradients up to $z=2.3$ \citep{miller22}.
These early findings are supported by several other works \citep[e.g.,][]{suess22, cutler24, ito24, ormerod24, martorano24, clausen25} that measured sizes of high-$z$ galaxies as a function of wavelength, and find that galaxies are systematically smaller in the near-IR than in the optical, consistent with the trends seen in the present-day universe. The early indications are that, at $z\approx 2$, dust gradients are more prominent than stellar population gradients, consistent with the notion of dust-obscured bulge growth at earlier cosmic times \citep{tadaki17, tacchella18, nelson19, tadaki20}. 
In fact, at high redshift, the highest $A_V$ values are seen among the most massive galaxies \citep[][]{price14,cullen18,lorenz24,nersesian25,van-der-wel25}, which have shallower optical profiles than in the near-IR \citep{martorano25}, suggesting that central mass concentrations -- bulges or forming bulges -- are present but strongly attenuated \citep[see also][]{benton24}.

Using the complementarity of HST and JWST observations, we extend the early JWST results to a significantly larger sample of galaxies ($>10,200$) with redshifts between 0.5 and 2.5 to investigate the origin and evolution across cosmic time of ${\textit{U}} - {\textit{V}}$ and ${\textit{V}} - {\textit{J}}$ color gradients. 
The two main goals are: (1) to determine the color gradients as a function of mass and galaxy type (star-forming vs.~quiescent), and (2) to interpret the cause of the gradients and its evolution with cosmic time.
The paper is structured as follows: in Section \ref{sec:Data} we present an overview of the sample selection (Sect. \ref{sec:sample}), how we construct light profiles across wavelength (Sect. \ref{sec:galfitting}), how we measure resolved rest-frame ${\textit{U}} - {\textit{V}}$ and ${\textit{V}} - {\textit{J}}$ colors and give an overview of the global galaxy properties (Sect. \ref{sec:rest-flux}); in Section \ref{sec:results} we present the results of this work: the color gradients (Sect. \ref{sec:color_grad}), their effect on the UVJ diagram (Sect. \ref{sec:UVJ}) and their relation with inclination as traced by the projected axis ratio (Sect. \ref{sec:col_q}). Finally, in Section \ref{sec:conclusion}, we summarize our results and draw our conclusions.
Throughout the paper we assume a standard Flat-$\Lambda$CDM cosmology with H$_0=70~$km$~$s$^{-1}~$Mpc$^{-1}$, $\Omega_m=0.3$ and we adopt the AB magnitude system \citep{oke83}.

\section{Data}\label{sec:Data}
    \subsection{Initial sample selection}\label{sec:sample}

    The parent sample is drawn from the Dawn JWST Archive (DJA) morphological catalog\footnote{\url{https://dawn-cph.github.io/dja/blog/2024/08/16/morphological-data/}}. This catalog contains photometric and morphological measurements for over $400,000$ galaxies in the five CANDELS \citep{grogin11, koekemoer11} fields observed with JWST/NIRCam over a plethora of different observational programs: e.g. PRIMER \citep{dunlop21}, COSMOS-Web \citep{casey23}, JADES \citep{eisenstein23}, FRESCO \citep{oesch23}, PANORAMIC \citep{williams25}, JEMS \citep{williams23}, and CEERS \citep{finkelstein17, Finkelstein23}. 

    Redshifts and global physical parameters ($M_\star$, rest-frame colors, $A_V$, which are relevant for this paper) are taken from the DJA morphological catalog as is. These were estimated via SED fitting running the code \textsc{Eazy} \citep{brammer08} on \SI{0.5}{\arcsecond} aperture photometry for all the available HST/ACS, HST/WFC3, JWST/NIRCam, and JWST/MIRI filters corrected to total fluxes. The high angular resolution and sensitivity of the four instruments on this wide wavelength range (from $0.4~\mu{\text{m}}$ to $21~\mu{\text{m}}$ though just for $\approx$ half targets JWST/MIRI photometry is available) make it possible to recover accurate redshift measurements as well as stellar-mass measurements. 
    \textsc{Eazy} was run on the \textit{agn$\_$blue$\_$sfhz$\_$13}\footnote{\url{https://github.com/gbrammer/eazy-photoz/tree/master/templates/sfhz}} template set, which consists of 13 templates from the Flexible Stellar Populations Synthesis code \citep[FSPS][]{conroy09,conroy10} built with a \cite{chabrier03} initial mass function (IMF) and a \cite{kriek13} dust attenuation law, a template derived from the NIRSpec spectrum of a $z=8.5$ galaxy presented by \cite{carnall23} and a template generated to replicate the JWST/NIRSpec spectrum of a $z=4.5$ source perhaps consistent with an obscured AGN torus \citep{killi24}. For all the details about the \textsc{Eazy} setup and the global photometry, we redirect the reader to \cite{valentino23}.

    From the parent sample, we select galaxies with an \textsc{Eazy} stellar mass $M_\star\geq10^{9.5}~\text{M}_\odot$. The low-mass, high-redshift galaxies selected via this mass threshold are at least 2 magnitudes brighter than the detection limit of the catalog in each field, implying that our sample is complete in stellar mass across the redshift range investigated here ($0.5<z<2.5$).
    An overview of the filters used throughout this work is presented in Table \ref{tab: filt_summary}. We do not use JWST/MIRI imaging data because of the lower spatial resolution compared to HST and JWST/NIRCam. This limits our analysis up to rest-frame $\sim1.3~\mu\text{m}$ at $z=2.5$ and up to $\sim3~\mu\text{m}$ at $z=0.5$.
    The mosaics of the COSMOS and UDS fields are split in half with an overlapping region leading to some duplicate objects in the catalog. For these objects we only retain the one furthest from the mosaic's edges.

    \begin{table}[]
        \centering
        \caption{Filters used throughout the work for S\'ersic profile fitting and to compute rest-frame colors.}
        {\tiny
        \begin{tabular}{c|c|c}
            COSMOS  & HST/ACS        & F435W F606W F814W \\
                    & HST/WFC3       & F125W F140W F160W \\
                    & JWST/NIRCam-SW & F090W F115W F150W F200W \\
                    & JWST/NIRCam-LW & F277W F356W F410M F444W \\
            \hline
            UDS     & HST/ACS        & F435W F606W F814W \\
                    & HST/WFC3       & F125W F140W F160W \\
                    & JWST/NIRCam-SW & F090W F115W F150W F200W \\
                    & JWST/NIRCam-LW & F277W F356W F410M F444W \\
            \hline
            GOODS-S & HST/ACS        & F435W F606W F775W F814W \\
                    & HST/WFC3       & F105W F125W F140W F160W \\
                    & JWST/NIRCam-SW & F090W F115W F150W F182M \\
                    &                & F200W F210M\\
                    & JWST/NIRCam-LW & F277W F335M F356W F410M \\
                    &                & F444W \\
            \hline
            GOODS-N & HST/ACS        & F435W F606W F775W F814W \\
                    & HST/WFC3       & F105W F125W F140W F160W \\
                    & JWST/NIRCam-SW & F090W F115W F150W F182M \\
                    &                & F200W F210M \\
                    & JWST/NIRCam-LW & F277W F335M F356W F410M \\
                    &                & F444W \\
        \end{tabular}
        }
        \label{tab: filt_summary}
    \end{table}

    \begin{figure}[h!]
        \centering
        \includegraphics[scale=0.37]{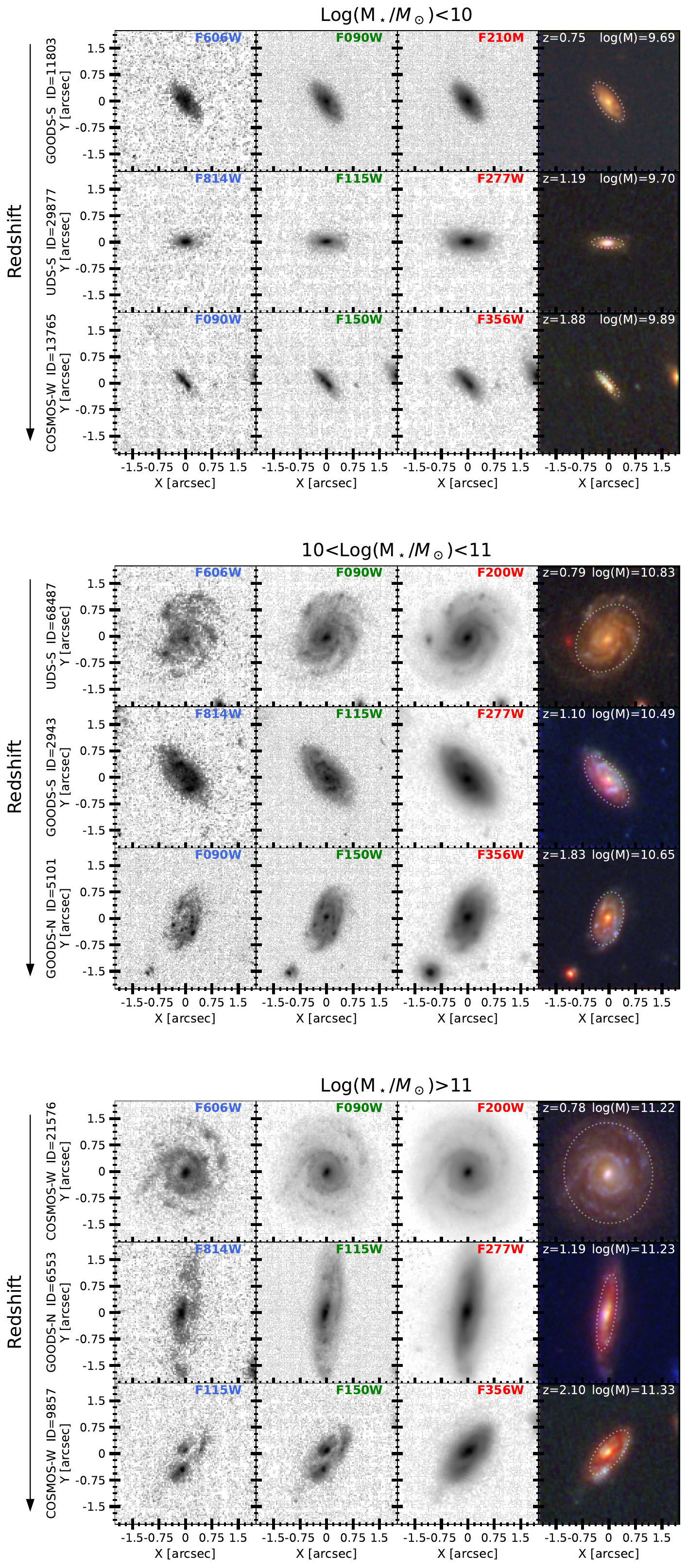}
        \caption{Example of 9 galaxies used in the work (field and DJA ID shown on the left of each row) divided into 3 sets according to their stellar mass and ordered by redshift. For each galaxy, we present cutouts in the three filters that closely match the rest-frame U(blue)-V(green)-J(red) bands. While the grey panels show the original images, to create the RGB images each filter was first deconvolved with a Richardson-Lucy algorithm. On the RGB image we overplot ellipses with semi-major axis twice the effective radii measured in JWST/NIRCam F444W as described in Sect. \ref{sec:galfitting}.
        }
        \label{fig:Gals}
    \end{figure}
    
    \begin{figure*}
                \centering
        	\includegraphics[scale=0.39]{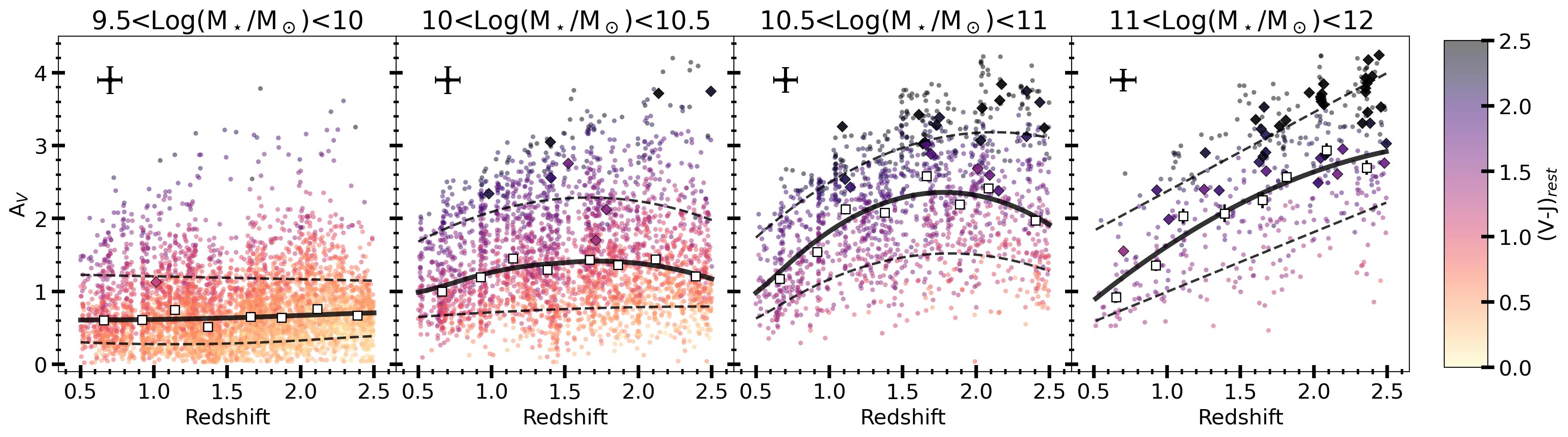}
                \caption{Attenuation in the \textit{V}-band ($A_V$) as a function of redshift in four stellar mass bins for star-forming galaxies color-coded with their ${\textit{V}} - {\textit{J}}$ color. White squares show the median in stellar mass bins and error bars the statistical uncertainty ($\sigma/\sqrt{N}$). The solid black lines show the spline-percentile regression while the dashed lines show the 16-84 percentiles of the distribution. Diamonds highlight sub-mm selected galaxies. 
                In the top left corner of each panel we show the median (16-84)/2 percentile ranges of the posterior distributions of $A_V$ and redshift as provided by \textsc{Eazy}. Massive galaxies at high redshift are redder and more attenuated than their lower redshift counterparts.}
            \label{fig:AV_z_mbins_ccVJ}
        \end{figure*}
    Our starting catalog includes $14,780$ galaxies. Figure \ref{fig:Gals} shows nine representative galaxies divided into three stellar-mass bins at three (increasingly higher) redshifts. For each galaxy, we present three cutouts in the filters closer to the rest-frame $U$, $V$, and $J$ bands, alongside their combination into an RGB image following a Richardson-Lucy deconvolution.
    
    In Figure \ref{fig:AV_z_mbins_ccVJ} we present an overview of the properties of the star-forming galaxies in our sample, showing the evolution of $A_V$ with redshift in four stellar mass bins. We classify galaxies as star-forming or quiescent as in \cite{muzzin13} based on UVJ colors.
    For star-forming galaxies with $M_\star~<~10^{10.5}~{\text{M}_\odot}$, $A_V$ shows little redshift dependence, specifically, $A_V$ is $\approx0.6$ for $M_\star<10^{10}~{\text{M}_\odot}$ and $\approx1.2$ for $10^{10}~<~M_\star/{\text{M}_\odot}~<~10^{10.5}$. More massive galaxies show a clear redshift evolution: from median $A_V\sim1$ at $z\sim0.5$ up to  $A_V\sim3$ at $z\sim2.5$. For a more detailed examination of the evolution of the colors and dust attenuation, we refer to \cite{van-der-wel25}.
        
    The existence of a population of dusty galaxies at $z\gtrsim 2$ has, of course, been known for decades, both from near-IR surveys \citep{franx03, daddi04, labbe05, wuyts07}, far infrared surveys \citep[e.g.,][]{shirley19}, and surveys at mm wavelengths \citep[e.g.,][]{smail97, smail02, chapman03, daddi07}. 
    To provide this context, we cross-match our catalog with the \textit{A$^3$COSMOS} and \textit{A$^3$GOODSS} catalogs \citep{adscheid24} that contain all galaxies identified by ALMA during all the programs that surveyed the COSMOS and GOODS-S fields. We find a match for 92 sub-mm galaxies with a separation below \SI{0.4}{\arcsecond} (these are indicated by diamond markers throughout the work). 
    These sub-mm galaxies are mostly seen at $z>1.5$ and tend to be more massive than $M_\star>10^{10.5}~{\text{M}_\odot}$. All are characterized by the reddest ${\textit{V}} - {\textit{J}}$ colors and have a median $A_V\sim3$. According to their UVJ colors and to the definition of quiescence adopted throughout this work \citep{muzzin13}, all these 92 galaxies are classified as star-forming. This sub-mm detected sample is not complete or representative for our sample; it is included to highlight the link between highly attenuated galaxies and dust emission-selected samples, as also done in \cite{van-der-wel25}.
    
    \subsection{S\'ersic Profile Fits}\label{sec:galfitting}
        To quantify color gradients, we use multi-wavelength S\'ersic profile fits. While these are crude, summary descriptions of the light distribution, they address two important problems. First, the profile fits account for the wavelength-dependent PSF effects, which flatten color gradients as seen directly in the images (see e.g., Fig. \ref{fig:Gals}). Second, pixel-to-pixel variance is averaged out. 
        To fairly compare color gradients between galaxies with a large variety of sizes and redshifts, we measure colors and gradients in apertures relative to the half-light radius using photometry based on the galaxies' 2D brightness model approximated with a single S\'ersic profile. We thus perform S\'ersic profile fitting of all the galaxies using the code \textsc{GalfitM} \citep{haussler13, vika13}, a further developed version of \textsc{Galfit} \citep{peng02,peng10}. The setup is overall the same as presented in \cite{martorano24} with minor differences. 
        
        Briefly, for each field we create a segmentation map using the code \textsc{SEP} \citep{bertin96, barbary16} using as detection image a stack of all long-wavelength JWST/NIRCam filters available (see Table \ref{tab: filt_summary} for the complete list of filters). We set a detection threshold at 3$\sigma$, a minimum area of 5 pixels and a contrast ratio for object deblending of 0.02. For each source in the parent sample, we create cutouts large enough to contain 25 times more pixels available for background estimation than pixels falling in a segment. We set a minimum size of the cutout of \SI{8}{\arcsecond} and a maximum of \SI{20}{\arcsecond}.
        Following \cite{martorano23} we fit simultaneously with the main source all objects identified in the segmentation map that are brighter (or up to one magnitude fainter) than the target. This ensures that the light of bright objects in the cutouts is properly subtracted, hence allowing for a better estimate of the background. Any other source in the cutout is masked with the corresponding segment of the segmentation map. The background is left as a free parameter in the fit performed by \textsc{GalfitM}.
        For each galaxy, we provide \textsc{GalfitM} with initial estimates for the position, magnitude, effective radius, and axis ratio based on the results obtained from the \textsc{SEP} analysis, while the initial guess for the position angle is set to 0 and for the S\'ersic index to 2.3.
        
        For each filter we retrieve a Point Spread Function (PSF) using the same procedure outlined in \cite{martorano24}: we select a sample of candidate stars from the size-brightness relation in the filter F444W, and observed in all the available filters; we masked all sources within \SI{3}{\arcsecond} and rejected those candidate stars with nearby bright objects which could potentially bias the background. From this set of objects, we visually select the most suitable stellar candidates: isolated, bright, but not saturated in the center. The selected stars are stacked after normalizing their \SI{0.3}{\arcsecond} aperture flux and background subtraction. For JWST/NIRCam, we match the flux of the stacked PSF in an annulus between \SI{2.5}{\arcsecond} and \SI{3}{\arcsecond} to the flux in the same annulus of the model PSF computed with \textsc{WebbPSF} \citep{perrin14}. This guarantees that any low-level residual background in the stacked PSF is removed.
        The PSFs obtained are used as input for \textsc{GalfitM}.    
        
        Throughout this work, we assume as reference filter F444W. We discard all galaxies whose fit in this filter failed or did not reach convergence for one of the parameters ($\sim 4\%$ of the whole sample).
        M.M. visually inspected the fit results for all the galaxies, flagging and removing evident mergers, gravitational lenses, and those where the flux of bright sources near the target strongly pollutes the photometry of the target itself. In addition, objects defined as segments including multiple bright peaks are removed.   Finally, 22 galaxies are removed whose S\'ersic index value in F444W hit the fit boundaries of 0.2-10. In total, another $5\%$ of the sample are thus rejected, leaving $13,332$ galaxies for which useful color gradients can be measured.
        The rejected targets are evenly distributed in stellar mass and redshift hence their removal does not severely bias our sample.

        For each galaxy, we compute rest-frame $0.5~\mu\text{m}$ and $1.5~\mu\text{m}$ effective radii ($R_{\text{e}, 0.5\,\mu{\text{m}}}$ and $R_{\text{e}, 1.5\,\mu{\text{m}}}$, respectively) fitting a second-order Chebyshev polynomial to all the effective radii calculated in the filters investigated as a function of wavelength. We iterate this procedure 1000 times Gaussian sampling the $R_\text{e}$ measurements from their uncertainty distributions. We consider the best polynomial the one made of the median of the coefficients retrieved. The same procedure is repeated to compute the S\'ersic index at rest-frame $0.5~\mu\text{m}$ ($n_{0.5\mu\text{m}}$) and $1.5~\mu\text{m}$ ($n_{1.5\mu\text{m}}$).
    
    \subsection{${\textit{U}} - {\textit{V}}$ and ${\textit{V}} - {\textit{J}}$ Rest-Frame Color Gradients}\label{sec:rest-flux}
    
        Using the parameters recovered from the Sérsic profile fitting presented in Section \ref{sec:galfitting}, we construct images of the intrinsic Sérsic profile (not PSF-convolved) in each filter.
        We measure fluxes within an ellipse whose center, semi-major axis, position angle, and axis ratio are taken from the F444W S\'ersic profile fit.
        Likewise, we measure aperture photometry in an elliptical annulus between 1 and 2 $R_{\text{e}, F444W}$.\footnote{$R_{\text{e}, F444W}$ traces different rest-frame wavelengths at different redshift. Using the rest frame aperture $R_{\text{e}, 1.5\,\mu{\text{m}}}$ does not change the results presented in this work.} We do not add flux residuals as done by, e.g., \cite{szomoru10} and \cite{suess19b} to account for imperfections in the single S\'ersic profile fits. The benefits of this approach do not, for our purposes, outweigh the additional uncertainties (noise). To verify that this is the case, we measure the residual F150W$-$F115W and F356W$-$F150W colors in the two apertures and their effect on the measured color gradient from the S\'ersic profiles. These filters cover rest-frame U, V, and J at $z\approx 2$. The average change in the colors (and gradients) is $<1\%$ and the scatter just $3\%$, which implies that the simplified description of the light profiles with single S\'ersic profiles does not lead to relevant systematic effects in the measured color gradients (see Appendix \ref{appendix:residuals}).

        We use \textsc{Eazy} to derive rest-frame \textit{U}, \textit{V} and \textit{J} fluxes by fitting the SED of both apertures, fixing the redshift at the value from the DJA catalog, and also adopting the same setup and zero-points as those used for the DJA catalog ensuring this does not induce systematic biases when comparing rest-frame fluxes retrieved in this work with those in the parent catalog.
        We define the color gradient as the color difference between the outer ($1-2\,R_{e, \text{F444W}}$) and inner ($<R_{e, \text{F444W}}$) apertures, adopting the convention that a galaxy with a red center and blue outskirts has a negative color gradient.

        The use of deconvolved S\'ersic profile models for the aperture photometry corrects for PSF effects that flatten gradients in the original, PSF-convolved images, especially for small galaxies. The drawback of this method is the reliance on the single-component S\'ersic models and the quality of the fits.

        To avoid the effects of template mismatch, we require that for each of the rest-frame \textit{U}, \textit{V}, \textit{J} bands at least $50\%$ overlaps with at least one of the HST or JWST filters listed in Table 1. That way, the calculated rest-frame colors minimize interpolation by approximating the rest-frame colors to observed ones, without strong template dependencies. $\approx 22\%$ of the galaxies are rejected by this constraint. Rejected galaxies are roughly uniformly distributed in redshift due to the field-to-field variation in filter sets, with the clear exception of lacking rest-frame $J$-band coverage at $0.88<z<0.90$ due to the large gap between F200W and F277W. 
        We note that the rejected $22\%$ do not strongly affect the median trends in the results below, but they would have caused an increase in scatter due to the larger uncertainties induced by the interpolation between filters by \textsc{Eazy}.

        The final sample contains $10,359$ galaxies, of which $9,444$ are star-forming and $915$ are quiescent. 71 out of the original 92 sub-mm selected galaxies survived the selection.

\section{Results} \label{sec:results}
    \begin{figure*}
            \centering
        \includegraphics[scale=0.65]{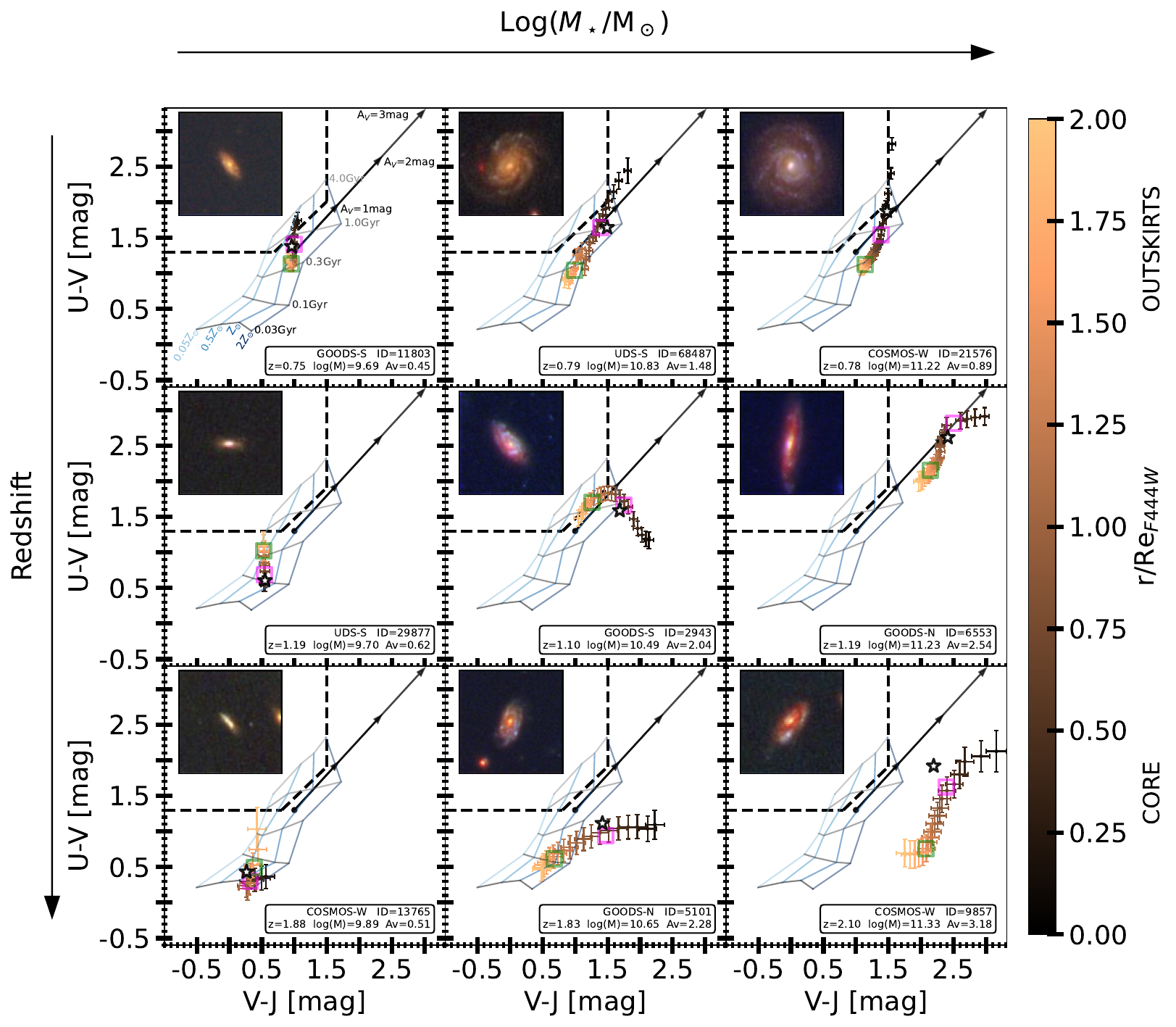}
            \caption{UVJ plots for the same galaxies presented in Figure \ref{fig:Gals}. Each panel shows the UVJ color gradients in elliptical apertures of width \SI{0.04}{\arcsecond} up to an aperture twice the effective radius in the filter F444W. Dashed black lines identify the quiescent region as defined in \cite{muzzin13}. The white star with black contour shows the UVJ colors of the galaxy as in the DJA catalog, while the magenta (green) squares show the color measured within $R_{\text{e}}$ ($1-2R_{\text{e}}$).
            In the bottom right corner we report the field and ID of the galaxy, together with its redshift, stellar mass and global Av. In each panel we show a grid of metallicities and ages from the \textsc{BPASS} stellar libraries together with the theoretical $A_V$ vector induced by a dust screen.
            This figure showcases different kind of color gradients across galaxies in the sample.
            }
        \label{fig:Grad_NiceGals}
    \end{figure*}
    The superb angular resolution of JWST in the rest-frame near-IR and extensive wavelength coverage provided by the synergy of HST and JWST allows us to probe beyond the global colors of galaxies, bringing into view the color gradients within them.
    Figure \ref{fig:Gals} already demonstrates a visually striking presence of color gradients in a small sample of galaxies. 
    The galaxies in the figure are representative of three stellar mass bins and three redshift bins. 
    While low-mass galaxies show similar colors and mild color gradients at any redshift, for $M_\star>10^{10.5}~{\text{M}_\odot}$, galaxies at $z>2$ are systematically redder than at $z<1$, and also have cores that are redder than the outskirts.
    For the same set of 9 galaxies shown in Figure \ref{fig:Gals}, in Figure \ref{fig:Grad_NiceGals} we present the ${\textit{U}} - {\textit{V}}$ and ${\textit{V}} - {\textit{J}}$ color gradients showing the variation of the colors up to $2~R_{e,F444W}$ on the UVJ plane. We also plot a grid of expected UVJ colors as a function of stellar metallicity and ages using the \textsc{BPASS} stellar libraries v.2.3.1 with solar $[\alpha/Fe]$ \citep{byrne22} together with the $A_V$ vector that represents the impact of a simple dust screen as modelled within the \textsc{THEMIS} modeling framework \citep{jones17}.
    The low-mass galaxies show just mild color gradients at any redshift while for higher stellar masses, color gradients are systematically stronger, with ${\textit{U}} - {\textit{V}}$ and ${\textit{V}} - {\textit{J}}$ changing by up to $1~\text{mag}$ between the core and the outskirts of the galaxy.
    The colors and color gradients for low mass galaxies seem compatible with a combination of age and metallicity gradients, while, at higher stellar masses, the need of dust (possibly with non-standard geometries) to redden the templates is clear.
    To support the qualitative color gradient trends shown in this Figure, in the next sections we quantify how the color gradients vary with global galaxy properties and redshift.
    
        \subsection{Color gradients}\label{sec:color_grad}

            \begin{figure*}
                \centering
            \includegraphics[scale=0.375]{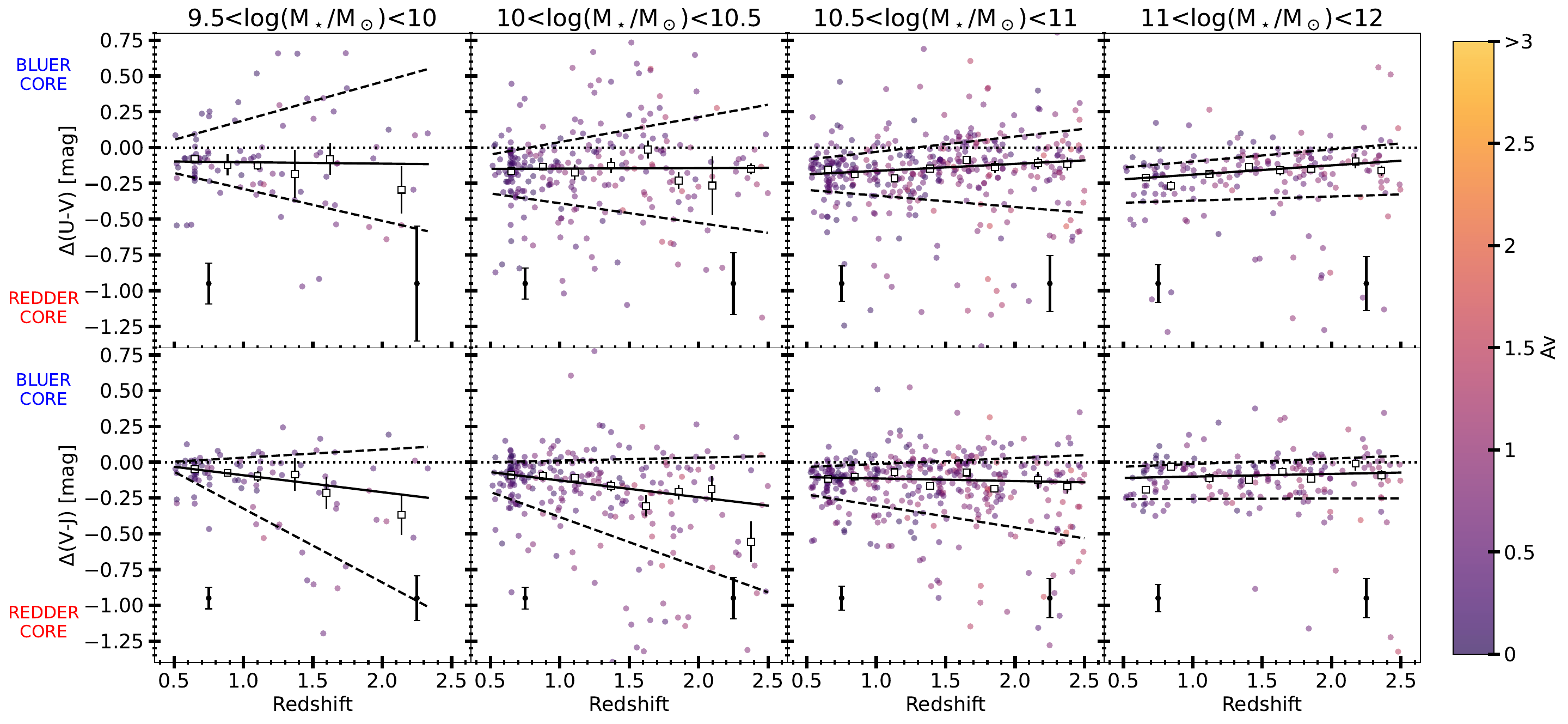}
                \caption{
                Color gradients as a function of redshift for quiescent galaxies in four stellar mass bins and color-coded with global $A_V$. The top row shows the ${\textit{U}} - {\textit{V}}$ color gradients ($\Delta(U-V)=(\text{U-V})_{1<r/R_\text{e}<2}-(\text{U-V})_{r<R_\text{e}}$) while the bottom row shows the ${\textit{V}} - {\textit{J}}$ color gradients ($\Delta(V-J)=(\text{V-J})_{1<r/R_\text{e}<2}-(\text{V-J})_{r<R_\text{e}}$). Black error bars in the lower left (right) corner of each panel show the median uncertainty on color gradients for all the galaxies in that stellar mass bin with $z<1.5$ ($z>1.5$).
                Solid black lines show the running median; dashed lines show the 16-84 percentile range. Squares show the median color in redshift bins of width 0.25 with (generally negligible) error bars representing the statistical uncertainty $\sigma/\sqrt{N}$.
                Both colors show a mild stellar mass dependence and a mild or absent redshift evolution. Just galaxies with $M_\star<10^{10.5}~{\text{M}_\odot}$ show a significant evolution of $\Delta(V-J)$ with redshift.
                }
                \label{fig:Grad_vs_massQ}
            \end{figure*}

            \begin{figure*}
                \centering
            	\includegraphics[scale=0.375]{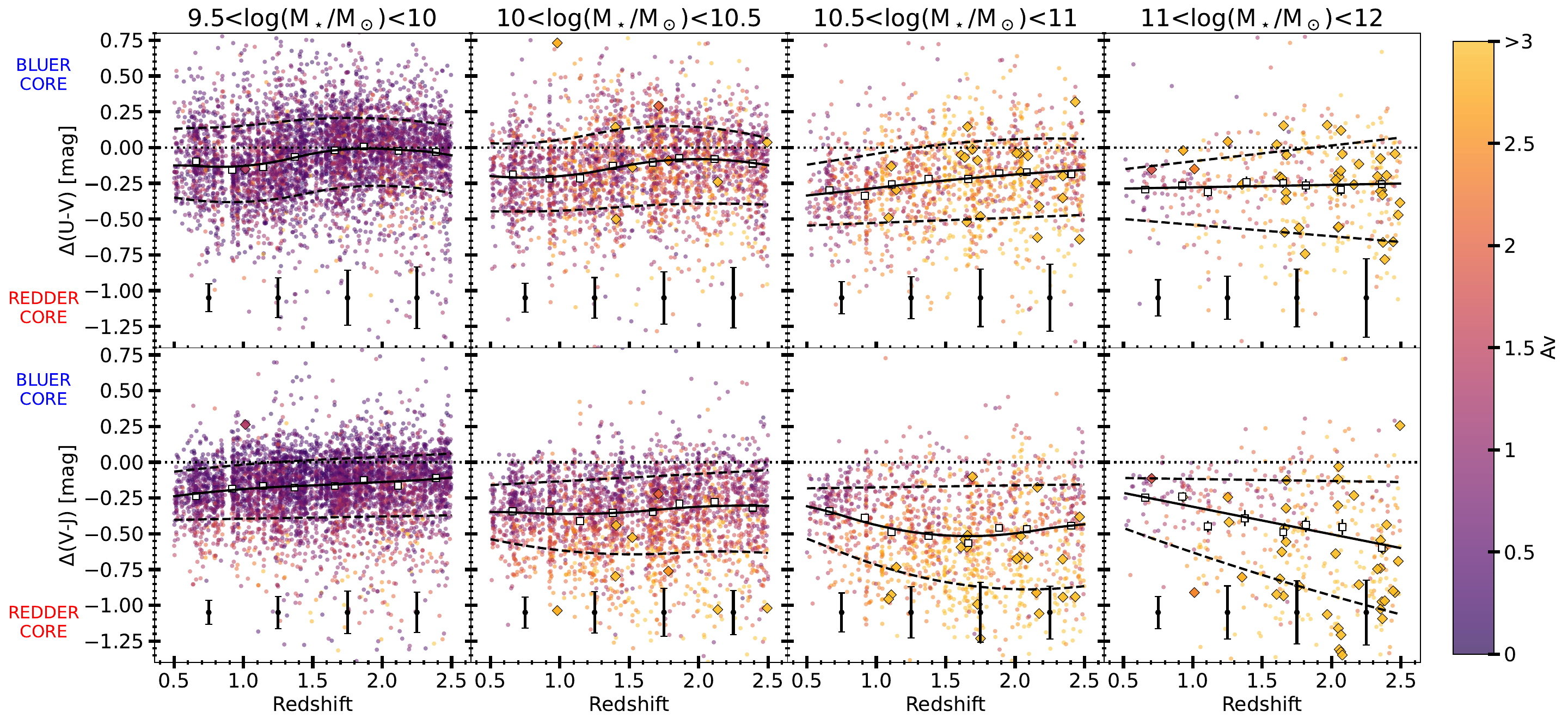}
                    \caption{Same as Fig. \ref{fig:Grad_vs_massQ} but for star-forming galaxies. Diamonds identify sub-mm selected galaxies. Black error bars in the lower part of each panel show the median uncertainty on color gradients at that redshift.
                    $\Delta(U-V)$ show no discernible redshift dependence and low-mass galaxies show no ${\textit{U}} - {\textit{V}}$ gradients at $z>1.5$. Conversely, $\Delta(V-J)$ shows a strong mass and redshift dependence.
                    }
                \label{fig:Grad_vs_massSF}
            \end{figure*}

            In Figures \ref{fig:Grad_vs_massQ} and \ref{fig:Grad_vs_massSF} we present the ${\textit{U}} - {\textit{V}}$ and ${\textit{V}} - {\textit{J}}$ color gradients (hereafter $\Delta(U-V)$ and $\Delta(V-J)$ respectively) of star-forming and quiescent galaxies as a function of redshift and in four different stellar mass bins. In the lower part of each panel, we show the median uncertainty on the color gradients (Appendix \ref{appendix:errors}). The solid black lines show the running median computed with the \textsc{cobs} library \citep{ng07,ng22}, which allows for a smoothed combination of a spline regression and quantile regression, highlighting non-linear trends.

            Figure \ref{fig:Grad_vs_massQ} shows that, on average, quiescent galaxies display mild ${\textit{U}} - {\textit{V}}$ gradients, with $\Delta(U-V)\approx-0.1~\text{mag}$, independent of redshift and mass. Their median $\Delta(V-J)$, in contrast, shows a mild redshift dependence: galaxies at $z>2$ and $M_\star<10^{10.5}~{\text{M}_\odot}$ show stronger gradients than their lower redshift counterpart, whereas this redshift evolution weakens and disappears at higher stellar mass (Spearman coefficients: for $M_\star<10^{10.5}~{\text{M}_\odot}$ $\rho=-0.26$ while for $M_\star>10^{10.5}~{\text{M}_\odot}$  $\rho=-0.06$). As highlighted by the dashed lines showing the 16-84 percentiles, for low mass galaxies the scatter in the color gradients strongly increases with redshift. This increased scatter is mostly due to the larger uncertainties in the color gradients (as shown by the systematically larger error bars in the bottom right corner), and suffers from small number statistics. We visually inspected galaxies with gradients stronger than the $16^{\text{th}}$ percentile (outliers) and found that, in most cases, the gradients are highly uncertain, with a small subset of objects with genuinely strong gradients.
            Conversely, for massive quiescent galaxies, the uncertainty in the color gradients exhibits only mild redshift dependence, reflecting the relative ease of measurement in larger objects \citep[see, e.g.,][]{martorano24} and the absence of pronounced galaxy-to-galaxy variations. The low number of low-mass quiescent galaxies at high redshift, combined with the larger uncertainties in their color gradients, reduces our confidence in claiming a physical trend with redshift for these objects despite the values retrieved for the Spearman coefficient.
            The global $A_V$ of quiescent galaxies also varies mildly with redshift, increasing up to $A_V\approx1$ \citep{van-der-wel25} as can be expected from their higher sSFR at high-$z$ \citep[e.g.,][]{leja22}. However, the $A_V$ estimates of such systems might be questionable given the systematic uncertainties related to our limited knowledge of the intrinsic near-IR colors of evolved stellar populations. Nevertheless, color gradients in quiescent galaxies primarily appear to originate with gradients in their stellar population (as also suggested by the \textsc{BPASS} comparison in Figure \ref{fig:Grad_NiceGals}) rather than by strong dust gradients \citep[see also e.g.,][]{suess20}.

            The conclusions change for star-forming galaxies (Fig. \ref{fig:Grad_vs_massSF}). These are characterized by generally negative $\Delta(U-V)$, except at the low-mass end ($M_\star<10^{10}~\text{M}_\odot$), where star-forming galaxies show no ${\textit{U}} - {\textit{V}}$ gradients at $z>1.5$, with the scatter dominated (as for quiescent galaxies) by random uncertainties due to limited sensitivity in the rest-frame $U$ band. Higher-mass galaxies have slightly stronger $\Delta(U-V)$, though with no or only weak redshift dependence. We find that $\Delta(U-V)$ is uncorrelated with global $A_V$ (Spearman rank correlation coefficient $\approx 0.1$). The $\Delta(U-V)$ gradients are broadly comparable with the HST-based ${\textit{U}} - {\textit{V}}$ gradients presented in \cite{wuyts20} despite the different methodology and photometry (they investigate color gradients measuring $\Delta(U-V)$ in apertures smaller and larger than $2~\text{kpc}$).
            
            The ${\textit{V}} - {\textit{J}}$ gradient $\Delta(V-J)$ shows more variation than  $\Delta(U-V)$, as well as stronger and more interesting trends with stellar mass and redshift. The uncertainties are small, even at large $z$, thanks to the exquisite depth of the NIRCam imaging. Gradients are significantly negative in general, even for low-mass galaxies \citep[see also][where $g-r$ gradients are investigated]{miller23}. There is a strong mass-dependence and a strong redshift dependence for higher-mass galaxies ($M_\star>10^{10.5}~\text{M}_\odot$): the median and scatter are very large at $z>1.5$. $\Delta(V-J)$ correlates strongly with global $A_V$ (Spearman rank correlation coefficient of $\approx -0.6$).
            The emergence of large fractions of high-$A_V$ galaxies at $z>1.5$ discussed above goes hand in hand with the emergence of strong ${\textit{V}} - {\textit{J}}$ color gradients. Among galaxies with $A_V>2$, gradients of $\Delta(V-J)\approx -1$~mag are not uncommon: these galaxies are 1 magnitude redder within their effective radius than outside. 
            The large variation among massive star-forming galaxies at high redshift is not (entirely) due to uncertainties, but reflects a true galaxy-to-galaxy difference in dust properties and viewing angle. This is further addressed in Section \ref{sec:col_q}.
            For reference, the population of 71 sub-mm selected galaxies in our sample have a median $A_V=3.1_{-0.7}^{+0.6}$, and $\Delta(U-V)=-0.2^{+0.2}_{-0.3}~\text{mag}$ and $\Delta(V-J)=-0.7^{+0.4}_{-0.3}~\text{mag}$, comparable to the general population of high-$A_V$ star-forming galaxies, suggesting a link between IR-bright galaxies and centrally concentrated dust distributions. 
            
            Contrary to present-day massive galaxies, in which color gradients are driven by a combination of star-formation, age, and metallicity gradients \citep{tortora10,tortora11, la-barbera09}, our results imply that for $M_\star>10^{10.5}~{\text{M}_\odot}$ the main driver of the color gradients in $z>1$ galaxies is attenuation by dust. 
            Several authors using other colors \citep[i.e., ][]{liu16, liu17, nelson16, wang17} or methodologies \citep{miller22, miller23, van-der-wel24, martorano24} reached the same conclusion. Strong color gradients, together with a characteristically high $A_V$ in massive star-forming galaxies at $z>1$ imply that these galaxies are building (or have already built) a core through intensive attenuated star-formation that is unveiled when seen in the near-IR \citep{nelson16, nelson19,miller22,le-bail24, benton24,tan24,nedkova24a, martorano25, maheson25}.
            Evidence for bulge building in heavily dust obscured galaxy cores at $z\sim2$ has also been revealed using high-resolution ALMA $870~\mu\text{m}$ imaging \citep[e.g.,][]{hodge16, tadaki17, tadaki20}.
            In Appendix \ref{sec:sizes} we compare the color gradients computed in this paper with the corresponding size ratio at different wavelengths. While these quantities both arise from changes in the light profile with wavelength, they are not identical.
            
            \subsection{Color gradients in the UVJ diagram} \label{sec:UVJ}
            
                \begin{figure*}
                        \centering
                	\includegraphics[scale=0.43]{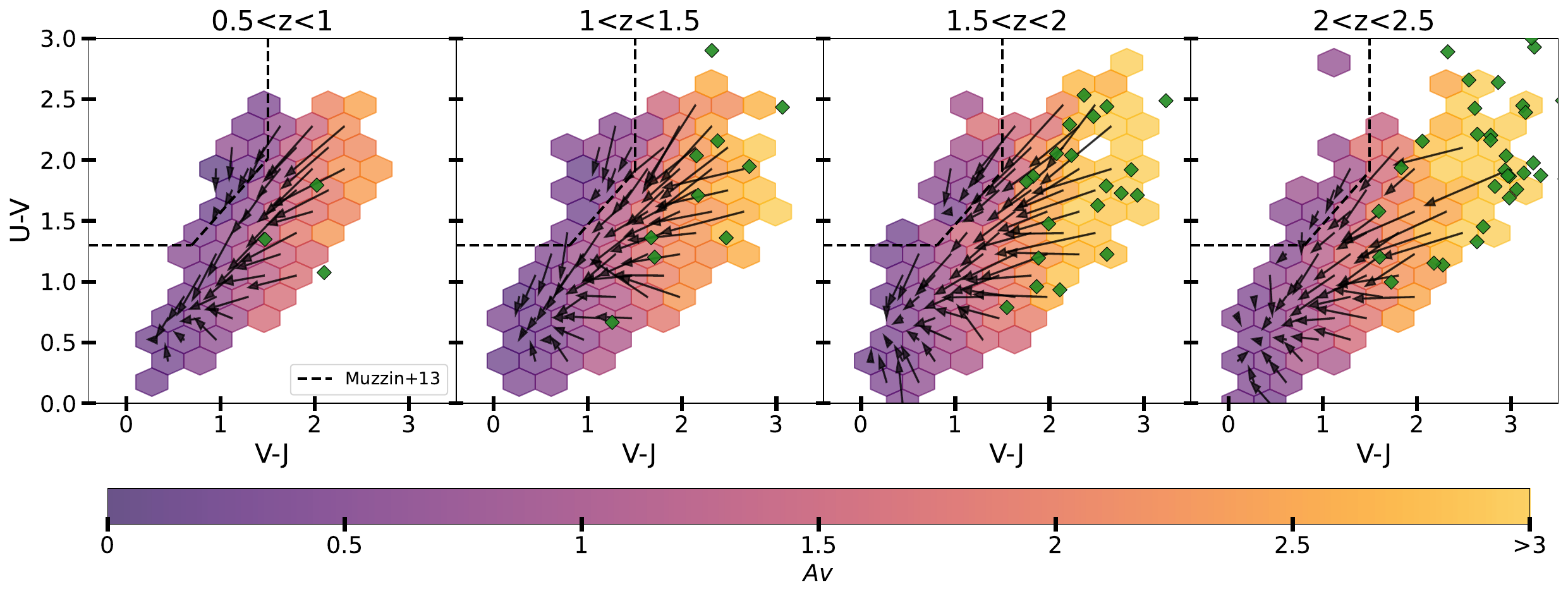}
                        \caption{UVJ diagram of colors measured within the F444W effective radius, with hexbins containing at least five galaxies color-coded with the median global $A_V$.
                        The length and direction of the arrows indicate the median color shift from within the effective radius to the aperture between 1$R_\text{e}$ and 2$R_\text{e}$, following our definition of color gradient throughout this paper. A minimum of 10 galaxies in the hexbin was required for drawing an arrow. Green diamonds highlight the ${\textit{U}} - {\textit{V}}$ and ${\textit{V}} - {\textit{J}}$ colors within 1$R_\text{e}$ for the sub-mm selected sample.
                        }
                    \label{fig:UVJ}
                \end{figure*}

            The color gradients presented in section \ref{sec:color_grad} have a major impact on the UVJ diagram, frequently used (as in this paper) to separate star-forming and quiescent galaxies. In Figure \ref{fig:UVJ}, we present the UVJ diagram using the rest-frame ${\textit{U}} - {\textit{V}}$ and ${\textit{V}} - {\textit{J}}$ colors within the effective radius, computed from elliptical aperture photometry with semi-major axis equal to $R_{e,F444W}$. 
            
            Similarly to global colors (see Figure \ref{fig:AV_z_mbins_ccVJ} and \citealt{van-der-wel25}), the central colors, computed within 1$R_\text{e}$, strongly correlate with the global $A_V$, with highly attenuated galaxies showing the reddest ${\textit{U}} - {\textit{V}}$ and ${\textit{V}} - {\textit{J}}$ colors. Once again, we see the emergence of a significant population of galaxies with very red centers (${\textit{V}} - {\textit{J}}\approx3$) at $z>1$, but without a noticeable reddening in ${\textit{U}} - {\textit{V}}$. 
            The color gradients, indicated by the black arrows in Figure \ref{fig:UVJ}, move the colors toward the bottom left corner of the UVJ plane, as expected for attenuation. 
            Galaxies with the bluest central ${\textit{U}} - {\textit{V}}$ colors (bottom left corner of each panel) show positive $\Delta(U-V)$ and mild or absent $\Delta(V-J)$, suggesting they are undergoing a central starburst. Conversely, star-forming galaxies with ${\textit{V}} - {\textit{J}}$~$<1~\text{mag}$ and ${\textit{U}} - {\textit{V}}$~$>0.6~\text{mag}$ show $\Delta(U-V)<0$ and $\Delta(V-J)\approx0$, perhaps a sign of inside-out quenching \citep{tacchella16} or of post-starburst systems \citep[e.g.,][]{belli19}.

            As we already witnessed in Figures \ref{fig:Grad_vs_massQ} and \ref{fig:Grad_vs_massSF}, quiescent galaxies are characterized by relatively mild color gradients.
            For $z<1$ (leftmost panel of Fig. \ref{fig:UVJ}) there are as many galaxies with quiescent centers and star-forming outskirts as vice versa ($\approx35\%$ of the quiescent population with $z<1$), and their median gradients are very small, putting both centers and outskirts within or near the quiescent boundary.
            Conversely, for $z>1$, there are $\sim3$ times more galaxies with star-forming centers and quiescent outskirts than the other way around. Those with star-forming center have relatively weak $\Delta(U-V)=-0.1^{+0.3}_{-0.3}~\text{mag}$ but strong $\Delta(V-J)=-0.4^{+0.3}_{-0.4}~\text{mag}$ (e.g., almost horizontal arrows in Fig. \ref{fig:UVJ}). The colors are too red to be explained by metallicity gradients, but rather require a grey attenuation curve in the UV-optical, which is indeed seen for more attenuated galaxies \citep{salim18, barisic20}. The high $A_V$ value of galaxies with near-horizonal gradient arrows in Fig. \ref{fig:UVJ} suggest that dust gradients are the likely explanation of the color gradients. Under this assumption, these galaxies have obscured, star-forming cores and quiescent, less attenuated outskirts.
            In the small subset of galaxies with quiescent cores and star-forming outskirts, the pattern is consistent with that observed in some present-day massive galaxies such as M31. These exhibit strong $\Delta(U-V)=-0.4^{+0.2}_{-0.5}~\text{mag}$ but weak $\Delta(V-J)=0.0^{+0.1}_{-0.2}~\text{mag}$ (see e.g., Fig. \ref{fig:Grad_NiceGals} top left panel). This may imply a genuine population gradient \citep[i.e.,][and references therein]{ellison18} like M31, or at least a gradient in star-formation activity, which affects ${\textit{U}} - {\textit{V}}$ more than ${\textit{V}} - {\textit{J}}$ \citep[e.g.,][]{gebek25}. Truly quiescent cores such as in M31 are relatively rare. Instead, it is more common to have star formation throughout, such as in local spiral galaxies like  M51, M83, M101 and the Milky Way. The color gradients are generally caused by a combination of factors, varying from galaxy to galaxy.
            Cosmological simulations are the ideal framework for properly identifying the factors responsible for global colors and color gradients.  At present, however, these still have problems describing all the features in observed UVJ diagrams \citep{donnari19, akins22, gebek25}. Following Gebek et al., however, we might expect our results to imply a specialized dust geometry; those authors find that obtaining large ${\textit{V}} - {\textit{J}}$ colors without strongly affecting the distribution of ${\textit{U}} - {\textit{V}}$ colors requires birthcloud-like attenuation around all stars younger than 1Gyr.

            The sub-mm selected galaxies, which are among the most attenuated of all galaxies in our sample, lay almost all in the top right corner of the UVJ diagram and have similar color gradients to all other galaxies with $A_V>3$.
            Only four out of the 71 sub-mm galaxies show color gradients resembling a star-forming core and quiescent outskirts. Two of them show extended disks with spiral arms and hints of ongoing minor mergers. 
            The other two show no morphological features. 

            \subsection{Color gradients and axis ratios: trends with inclination} \label{sec:col_q}

            \begin{figure*}
                \centering
                \includegraphics[scale=0.39]{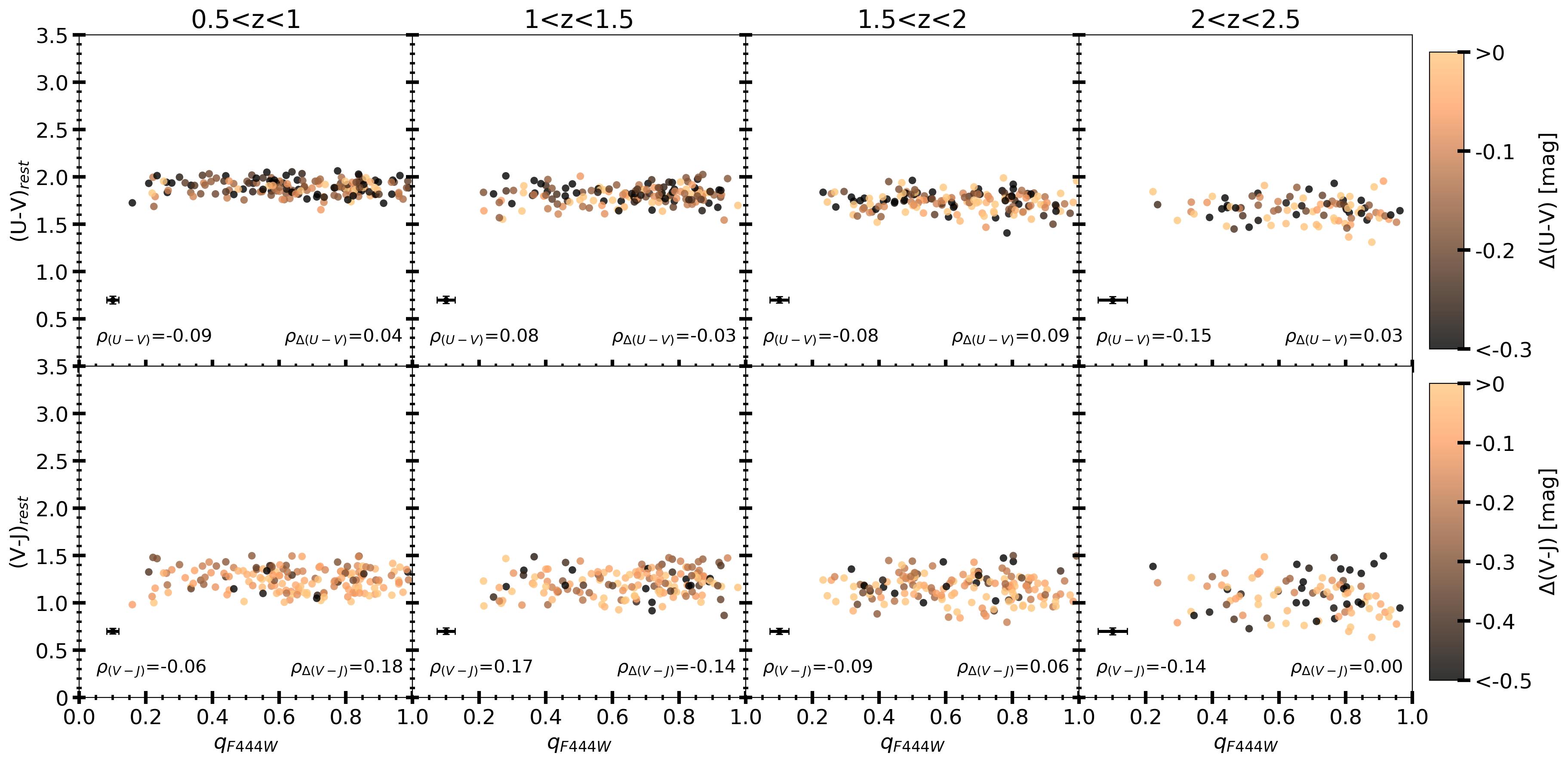}
                \caption{Projected axis ratio ($q$) observed in JWST/NIRCam-F444W against ${\textit{U}} - {\textit{V}}$ (upper panels) and ${\textit{V}} - {\textit{J}}$ (lower panels) colors in four redshift bins. Color-coding shows the color gradient computed as the difference between the color between 1-2$R_\text{e}$ and 0-1$R_\text{e}$. Just quiescent galaxies with stellar mass $M_\star\geq10^{10.5}~{\text{M}_\odot}$ are shown. 
                In the lower left corner, we show the median uncertainties on the rest frame colors and axis ratios and the Spearman correlation coefficient between the axis ratio and the color, while in the right corner, we show that between the axis ratio and the color gradient.
                }
                \label{fig:Col_q_Q}
            \end{figure*}
            
            \begin{figure*}
                    \centering
                    \includegraphics[scale=0.39]{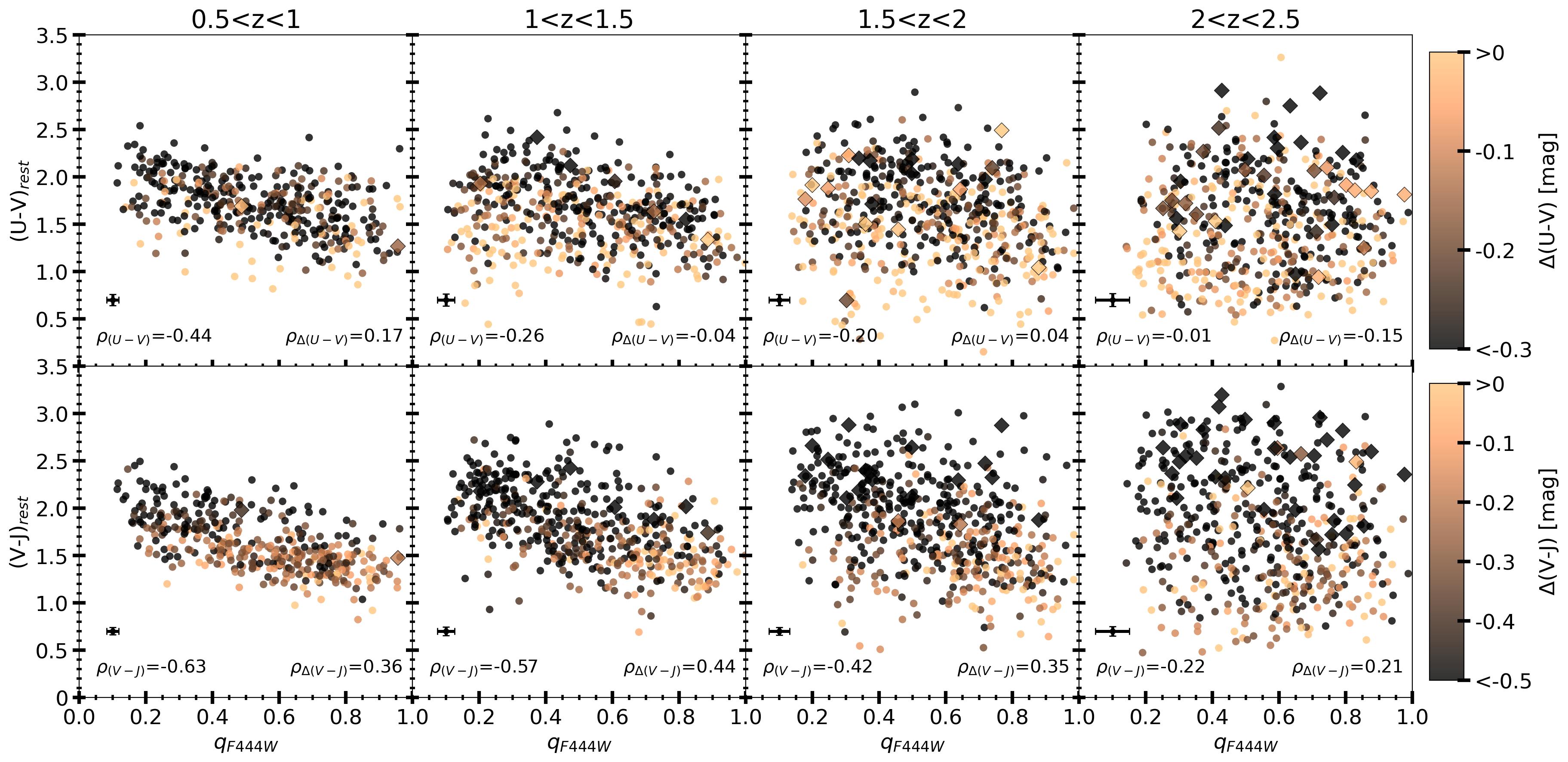}
                    \caption{Same as Figure \ref{fig:Col_q_Q} but for $M_\star\geq10^{10.5}~{\text{M}_\odot}$ star-forming galaxies. Diamonds identify sub-mm-selected galaxies.
                    The correlation between the colors and $q$ decreases with redshift.}
                    \label{fig:Col_q_SF}
                \end{figure*}
            
            The results presented in the previous sections indicate that, in massive star-forming galaxies at $z>1.5$, dust attenuation in the rest-frame $\textit{V}$-band is centrally concentrated within galaxies. At lower $z$, integrated (galaxy-averaged) $A_{V}$ values are lower and gradients weaker. To further explore the nature of this evolution with redshift we now consider the connection between axis ratio, as a proxy for viewing angle and attenuation properties (as traced by colors and color gradients) for quiescent and star-forming galaxies with $M_\star > 10^{10.5}~\text{M}_\odot$. The motivation for this mass selection is that those star-forming galaxies are most likely to be disk-like in both a geometric sense \citep[oblate and flattened][]{van-der-wel14a, zhang19} and as expressed by their kinematic properties \citep[e.g.,][]{forster-schreiber09, wisnioski15}.

            For disks (flat oblates), the projected axis ratio $q$, as measured in the F444W filter by the two-dimensional S\'ersic light profile fit, is a good tracer of the inclination, that is, a one-dimensional viewing angle. In Figure \ref{fig:Col_q_Q} and \ref{fig:Col_q_SF} we show the distribution of $q$ and the global ${\textit{U}} - {\textit{V}}$ and ${\textit{V}} - {\textit{J}}$ colors in bins of redshift for the quiescent and star-forming population respectively. 

            For the population of quiescent galaxies, little correlation between color and axis ratio $q$ is expected (as suggested by their tight color sequences, see \citealt{van-der-wel25}, and confirmed by the low Spearman correlation coefficients reported in the bottom left corner of each panel of Fig. \ref{fig:Col_q_Q}). Many, if not most, of these galaxies are shaped like oblate disks \citep{chang13a} up to at least $z=2$. The absence of a correlation with $q$ then implies that color is independent of inclination, which we interpret as evidence for transparency, that is, small dust columns. This is in line with the recent result that even quiescent galaxies with significant molecular gas reservoirs do not have much dust \citep{spilker25}. 
            Changing the definition of the quiescent boundary in the UVJ diagram adopted in this work -- now set to ${\textit{V}} - {\textit{J}}$~$<1.5$ \citep{muzzin13} -- does not alter the conclusions drawn on the basis of Fig. \ref{fig:Col_q_Q}. Further investigation of dust in quiescent galaxies requires a selection based on the sSFR (possibly retrieved from non-parametric SED fitting including far-IR photometry) to avoid any color-selection systematic.
             
            Conversely, for star-forming galaxies at $z<1.5$ the trend is familiar and expected: flat (edge-on) galaxies are redder than round (face-on) galaxies. The connection between color and viewing angle (well traced by the axis ratio for low-z star-forming galaxies, see e.g., \citealt{van-der-wel14a}) was previously established at $z\approx 1$ by \citep{patel12}. The small scatter in color at fixed $q$ (0.2 mag in ${\textit{V}} - {\textit{J}}$) is also noteworthy: variation among the galaxies in terms of their intrinsic properties (star-formation, stellar population, dust content) must be relatively small and to a large extent the observed colors are determined by inclination. We also see a correlation between color gradient and axis ratio: face-on galaxies (large $q$) have weaker gradients than edge-on (low $q$) galaxies (Spearman correlation coefficient $\rho_{(U-V)}\approx-0.3$ and  $\rho_{(V-J)}\approx-0.6$). In edge-on galaxies, the dust obscures the central region more than the outer parts. This correlation is weaker than the $q$-color correlation, but still significant, especially for $\Delta(V-J)$ (in fact, $\rho_{\Delta(U-V)}\approx0.1$ and $\rho_{\Delta(V-J)}\approx0.4$). Taking these trends together, we arrive at a picture that is consistent with the result that massive present-day spiral galaxies (with $V_c > 120$ km s$^{-1}$; $M_\star \gtrsim 10^{10}~\text{M}_\odot$) show thin, smooth, regular and galaxy-wide dust lanes aligned with the gravitationally dominant stellar disk \citep{dalcanton04}.

            But at $z>1.5$ the trends change and the correlations between $q$ and colors / color gradients weaken (with $\rho_{(U-V)}$ that drops to $\sim-0.1$ and $\rho_{(V-J)}$ to $\approx-0.2$). Viewing angle no longer matters in the same manner as at later cosmic times. To some extent, this is due to the less obviously disk-like nature of at least a subset of the galaxies. Rotational support decreases with redshift \citep{kassin12, wisnioski15, ubler19}, and the distribution of geometries is more varied \citep{van-der-wel14a, zhang19} so that axis ratio no longer tracks viewing angle in a unique manner. But the disappearance of disks is not the full story: a large fraction of galaxies is still disk-like in nature, the evidence for which was recently bolstered by the large observed fraction of galaxies with spiral arms in this mass and redshift range \citep[e.g.,][found $72\%$ of the galaxies with $2<z<2.5$ to be disk-like and $20\%$ to host spiral arms]{espejo-salcedo25b}, the existence of which are only possible in relatively thin stellar and/or gaseous disks.

            Yet the fact remains that the correlations between $q$ and the color (gradient) information are less obvious in statistical terms and less straightforward to interpret. At $z>2$ the $q - (V-J)$ correlation is still significant ($\rho_{(V-J)}=-0.22$), but weaker than at $z<1.5$, and with a very clear increase in scatter at fixed $q$ (0.7 mag at $z>2$ compared to 0.2 mag at $z<1$).  Likewise, the $q$-$\Delta(V-J)$ correlation persists at $z>2$ ($\rho_{\Delta(V-J)}=0.21$), if slightly weaker, with stronger gradients for flat (edge-on) galaxies. For ${\textit{U}} - {\textit{V}}$ the picture is different from ${\textit{V}} - {\textit{J}}$.

            At $z>2$ there is no $q - (U-V)$ correlation ($\rho_{(U-V)}\approx0$), and, if anything, a weak anti-correlation between $q$ and $\Delta(U-V)$ ($\rho_{\Delta(U-V)}=-0.15$), with slightly stronger gradients in round (face-on) galaxies. This trend is driven by a population of galaxies with blue integrated ${\textit{U}} - {\textit{V}}$ colors, weak ${\textit{U}} - {\textit{V}}$ gradients, but red ${\textit{V}} - {\textit{J}}$ colors and strong ${\textit{V}} - {\textit{J}}$ gradients. This population is already apparent at $1<z<1.5$, and a prototypical example is shown in the center of the 3$\times$3 set of panels in Figure \ref{fig:Grad_NiceGals}. The ${\textit{U}} - {\textit{V}}$ trends are explained by a patchy but not clearly radially varying distribution of relatively unobscured star-formation across the side of the underlying disk that is pointed in the direction of the observer.

            Taken together, these color and color gradient trends with $q$ clearly show that the dust-to-star geometry at $z>1.5$ is fundamentally different from the thin, regular dust-lane geometry seen at $0<z<1.5$. The evidence suggests that the dust distribution still retains some of the axisymmetric characteristics of a dust lane, given the bluer colors of face-on galaxies. But the much larger $A_{V}$ values also imply that the dust lane must be more vertically extended, looking more like a patchy thick disk, obscuring the majority of all intrinsic rest-frame ${\textit{V}}$-band light. Such a thick disk of dust must also be centrally concentrated, as implied by the color gradients, which are stronger for edge-on, high-$A_{V}$ galaxies (see correlation coefficients in Figure \ref{fig:Col_q_SF}). Finally, the dust distribution is also less smooth and regular than their thin present-day counterparts, as indicated by the large scatter in the ${\textit{U}} - {\textit{V}}$ and ${\textit{V}} - {\textit{J}}$ colors and their gradients. This dust model closely matches the one discussed in the rightmost panel of Fig. 7 of \cite{gebek25}.

\section{Conclusions} \label{sec:conclusion}

In this paper we used the combination of HST and JWST imaging to quantify and examine gradients in rest frame ${\textit{U}} - {\textit{V}}$ and ${\textit{V}} - {\textit{J}}$ colors for over $10,200$ galaxies at redshifts $0.5<z<2.5$ with stellar masses $M_\star>10^{9.5}~{\text{M}}_\odot$. 
To stabilize our color measurements against issues related to background subtraction and the PSF, we perform multi-wavelength 2D brightness modeling that fits a S\'ersic profile to each galaxy at each observed wavelength. We then measure the colors from these S\'ersic profiles.

${\textit{U}} - {\textit{V}}$ color gradients in star-forming galaxies (Fig. \ref{fig:Grad_vs_massSF}) show a mild stellar mass dependence -- with stronger gradients for more massive galaxies -- and a weak redshift evolution.
Their ${\textit{V}} - {\textit{J}}$ color gradients, on the other hand, are strongly mass and redshift dependent, with high-mass $2<z<2.5$ galaxies showing the strongest gradients on average. 
The central regions, in particular, have much redder ${\textit{V}} - {\textit{J}}$ colors than can be explained by any dust-free stellar population. This, in combination with the strong link between ${\textit{V}} - {\textit{J}}$ gradient and global $A_V$ we find, suggests that the optical and near-IR colors of high mass galaxies reflect a strong attenuation gradient and highly obscured centers.
Together with stellar population gradients, strong dust gradients produce ${\textit{U}} - {\textit{V}}$ and ${\textit{V}} - {\textit{J}}$ gradients that systematically shift the galaxy population within the UVJ color plane. Finally, the strongest gradients and reddest colors are seen for edge-on galaxies at $z>1.5$.  The high $A_{V}$ values imply a vertically extended dust geometry in the plane of the stellar disk, rather than the thin dust lanes seen at lower $z$. 

Meanwhile, the quiescent galaxy population has negative radial gradients in both ${\textit{U}} - {\textit{V}}$ and ${\textit{V}} - {\textit{J}}$ colors, with little or no redshift evolution. The lack of a relation between viewing angle/inclination and color implies that these galaxies are largely transparent and, in line with our knowledge of gradients in present-day early-type galaxies, that the color gradients are the result of stellar population gradients. The lack of a redshift evolution in the strength of the gradient may be the result of weakening age-induced gradients with cosmic time that are compensated for by increasing metallicity gradients, but further study is needed to confirm this picture.

The results of this work serve as a benchmark for the examination of the origin of gradients in cosmological simulations \citep[e.g.,][]{donnari19, akins22, gebek25}, to help understand the dust geometry and structure of galaxies 10 Gyr ago.

\begin{acknowledgements}
      MM acknowledges the financial support of the Flemish Fund for Scientific Research (FWO-Vlaanderen), research project G030319N.
      (Some of) The data products presented herein were retrieved from the Dawn JWST Archive (DJA). DJA is an initiative of the Cosmic Dawn Center (DAWN), which is funded by the Danish National Research Foundation under grant DNRF140.
\end{acknowledgements}

\bibliographystyle{aa} 
\bibliography{mypapers.bib}

@article{benton24,
	adsurl = {https://ui.adsabs.harvard.edu/abs/2024ApJ...974L..28B},
	archiveprefix = {arXiv},
	author = {{Benton}, Chlo{\"e} E. and {Nelson}, Erica J. and {Miller}, Tim B. and {Bezanson}, Rachel and {Gibson}, Justus and {Hartley}, Abigail I. and {Martorano}, Marco and {Price}, Sedona H. and {Suess}, Katherine A. and {van der Wel}, Arjen and {van Dokkum}, Pieter and {Weaver}, John R. and {Whitaker}, Katherine E.},
	doi = {10.3847/2041-8213/ad7e27},
	eid = {L28},
	eprint = {2409.08328},
	journal = {\apjl},
	month = oct,
	number = {2},
	pages = {L28},
	primaryclass = {astro-ph.GA},
	title = {{JWST Reveals Bulge-dominated Star-forming Galaxies at Cosmic Noon}},
	volume = {974},
	year = 2024}

@article{clausen25,
	adsurl = {https://ui.adsabs.harvard.edu/abs/2025ApJ...993..106C},
	archiveprefix = {arXiv},
	author = {{Clausen}, Maike and {Momcheva}, Ivelina G. and {Whitaker}, Katherine E. and {Cutler}, Sam E. and {Bezanson}, Rachel S. and {Dunlop}, James S. and {Grogin}, Norman A. and {Koekemoer}, Anton M. and {McLeod}, Derek J. and {McLure}, Ross J. and {Miller}, Tim B. and {Nelson}, Erica J. and {van der Wel}, Arjen and {Wake}, David A. and {Wuyts}, Stijn},
	doi = {10.3847/1538-4357/ae03aa},
	eid = {106},
	eprint = {2501.04788},
	journal = {\apj},
	month = nov,
	number = {1},
	pages = {106},
	primaryclass = {astro-ph.GA},
	title = {{The Evolution of Half-mass Radii and Color Gradients for Young and Old Quiescent Galaxies at 0.5 < z < 3 with JWST/PRIMER}},
	volume = {993},
	year = 2025}

@article{maheson25,
	adsurl = {https://ui.adsabs.harvard.edu/abs/2025arXiv250415346M},
	archiveprefix = {arXiv},
	author = {{Maheson}, Gabriel and {Tacchella}, Sandro and {Belli}, Sirio and {Park}, Minjung and {Danhaive}, A. Lola and {Bugiani}, Letizia and {Davies}, Rebecca and {Emami}, Razieh and {Khoram}, Amir H. and {Lam}, Laurence and {Leja}, Joel and {Mendel}, Trevor and {Nelson}, Erica June},
	doi = {10.48550/arXiv.2504.15346},
	eid = {arXiv:2504.15346},
	eprint = {2504.15346},
	journal = {arXiv e-prints},
	month = apr,
	pages = {arXiv:2504.15346},
	primaryclass = {astro-ph.GA},
	title = {{Big, Dusty Galaxies in Blue Jay: Insights into the Relationship Between Morphology and Dust Attenuation at Cosmic Noon}},
	year = 2025}

@article{ren25,
	adsurl = {https://ui.adsabs.harvard.edu/abs/2025ApJ...982..200R},
	archiveprefix = {arXiv},
	author = {{Ren}, Jian and {Liu}, F.~S. and {Li}, Nan and {Zhao}, Pinsong and {Cui}, Qifan and {Song}, Qi and {Li}, Yubin and {Mo}, Hao and {Yesuf}, Hassen M. and {Wang}, Weichen and {An}, Fangxia and {Zheng}, Xian Zhong},
	doi = {10.3847/1538-4357/adb961},
	eid = {200},
	eprint = {2502.15569},
	journal = {\apj},
	month = apr,
	number = {2},
	pages = {200},
	primaryclass = {astro-ph.GA},
	title = {{The Evolution of the Size and Merger Fraction of Submillimeter Galaxies across 1 < z {\ensuremath{\lesssim}} 6 as Observed by JWST}},
	volume = {982},
	year = 2025}

@article{spilker25,
	adsurl = {https://ui.adsabs.harvard.edu/abs/2025ApJ...993L..40S},
	archiveprefix = {arXiv},
	author = {{Spilker}, Justin S. and {Whitaker}, Katherine E. and {Narayanan}, Desika and {Bezanson}, Rachel and {Bodansky}, Sarah and {D'Onofrio}, Vincenzo R. and {Feldmann}, Robert and {Goulding}, Andy D. and {Greene}, Jenny E. and {Kriek}, Mariska and {Luo}, Yuanze and {Setton}, David J. and {Suess}, Katherine A. and {van der Wel}, Arjen and {Verrico}, Margaret E. and {Williams}, Christina C. and {Woodrum}, Charity and {Wu}, Po-Feng},
	doi = {10.3847/2041-8213/ae14d6},
	eid = {L40},
	eprint = {2507.16914},
	journal = {\apjl},
	month = nov,
	number = {2},
	pages = {L40},
	primaryclass = {astro-ph.GA},
	title = {{Unusually High Gas-to-dust Ratios Observed in High-redshift Quiescent Galaxies}},
	volume = {993},
	year = 2025}

@article{van-der-wel25,
	adsurl = {https://ui.adsabs.harvard.edu/abs/2025A&A...701A..30V},
	archiveprefix = {arXiv},
	author = {{van der Wel}, A. and {Martorano}, M. and {Marchesini}, D. and {Wuyts}, S. and {Bell}, E.~F. and {Meidt}, S.~E. and {Gebek}, A. and {Brammer}, G.~B. and {Whitaker}, K.~E. and {Bezanson}, R. and {Nelson}, E.~J. and {Rudnick}, G.~H. and {Kriek}, M. and {Leja}, J. and {Dunlop}, J.~S. and {Casey}, C.~M. and {Kartaltepe}, J.~S.},
	doi = {10.1051/0004-6361/202555488},
	eid = {A30},
	eprint = {2506.23669},
	journal = {\aap},
	month = sep,
	pages = {A30},
	primaryclass = {astro-ph.GA},
	title = {{Even redder than we knew: Color and A$_{V}$ evolution up to z = 2.5 from JWST/NIRCam photometry}},
	volume = {701},
	year = 2025}

@article{williams25,
	adsurl = {https://ui.adsabs.harvard.edu/abs/2025ApJ...979..140W},
	archiveprefix = {arXiv},
	author = {{Williams}, Christina C. and {Oesch}, Pascal A. and {Weibel}, Andrea and {Brammer}, Gabriel and {Cloonan}, Aidan P. and {Whitaker}, Katherine E. and {Barrufet}, Laia and {Bezanson}, Rachel and {Bowler}, Rebecca A.~A. and {Dayal}, Pratika and {Franx}, Marijn and {Greene}, Jenny E. and {Hutter}, Anne and {Ji}, Zhiyuan and {Labb{\'e}}, Ivo and {Manning}, Sinclaire M. and {Maseda}, Michael V. and {Xiao}, Mengyuan},
	doi = {10.3847/1538-4357/ad97bc},
	eid = {140},
	eprint = {2410.01875},
	journal = {\apj},
	month = feb,
	number = {2},
	pages = {140},
	primaryclass = {astro-ph.GA},
	title = {{The PANORAMIC Survey: Pure Parallel Wide Area Legacy Imaging with JWST/NIRCam}},
	volume = {979},
	year = 2025}

@article{espejo-salcedo25b,
	adsurl = {https://ui.adsabs.harvard.edu/abs/2025A&A...700A..42E},
	archiveprefix = {arXiv},
	author = {{Espejo Salcedo}, J.~M. and {Pastras}, S. and {V{\'a}cha}, J. and {Pulsoni}, C. and {Genzel}, R. and {F{\"o}rster Schreiber}, N.~M. and {Jolly}, J.-B. and {Barfety}, C. and {Chen}, J. and {Tozzi}, G. and {Liu}, D. and {Lee}, L.~L. and {Wuyts}, S. and {Tacconi}, L.~J. and {Davies}, R. and {{\"U}bler}, H. and {Lutz}, D. and {Wisnioski}, E. and {Shangguan}, J. and {Lee}, M. and {Price}, S.~H. and {Eisenhauer}, F. and {Renzini}, A. and {Nestor Shachar}, A. and {Herrera-Camus}, R.},
	doi = {10.1051/0004-6361/202554725},
	eid = {A42},
	eprint = {2503.21738},
	journal = {\aap},
	month = aug,
	pages = {A42},
	primaryclass = {astro-ph.GA},
	title = {{Galaxy morphologies at cosmic noon with JWST: A foundation for exploring gas transport with bars and spiral arms}},
	volume = {700},
	year = 2025}

@article{menanteau01,
	adsurl = {https://ui.adsabs.harvard.edu/abs/2001MNRAS.322....1M},
	archiveprefix = {arXiv},
	author = {{Menanteau}, F. and {Abraham}, R.~G. and {Ellis}, R.~S.},
	doi = {10.1046/j.1365-8711.2001.04028.x},
	eprint = {astro-ph/0007114},
	journal = {\mnras},
	month = mar,
	number = {1},
	pages = {1-12},
	primaryclass = {astro-ph},
	title = {{Evidence for evolving spheroidals in the Hubble Deep Fields North and South}},
	volume = {322},
	year = 2001}

@article{hinkley01,
	adsurl = {https://ui.adsabs.harvard.edu/abs/2001ApJ...560L..41H},
	archiveprefix = {arXiv},
	author = {{Hinkley}, Sasha and {Im}, Myungshin},
	doi = {10.1086/323940},
	eprint = {astro-ph/0109532},
	journal = {\apjl},
	month = oct,
	number = {1},
	pages = {L41-L44},
	primaryclass = {astro-ph},
	title = {{Optical-Near-Infrared Color Gradients in Early-Type Galaxies at z<=1.0}},
	volume = {560},
	year = 2001}

@article{ubler19,
	adsurl = {https://ui.adsabs.harvard.edu/abs/2019ApJ...880...48U},
	archiveprefix = {arXiv},
	author = {{{\"U}bler}, H. and {Genzel}, R. and {Wisnioski}, E. and {F{\"o}rster Schreiber}, N.~M. and {Shimizu}, T.~T. and {Price}, S.~H. and {Tacconi}, L.~J. and {Belli}, S. and {Wilman}, D.~J. and {Fossati}, M. and {Mendel}, J.~T. and {Davies}, R.~L. and {Beifiori}, A. and {Bender}, R. and {Brammer}, G.~B. and {Burkert}, A. and {Chan}, J. and {Davies}, R.~I. and {Fabricius}, M. and {Galametz}, A. and {Herrera-Camus}, R. and {Lang}, P. and {Lutz}, D. and {Momcheva}, I.~G. and {Naab}, T. and {Nelson}, E.~J. and {Saglia}, R.~P. and {Tadaki}, K. and {van Dokkum}, P.~G. and {Wuyts}, S.},
	doi = {10.3847/1538-4357/ab27cc},
	eid = {48},
	eprint = {1906.02737},
	journal = {\apj},
	month = jul,
	number = {1},
	pages = {48},
	primaryclass = {astro-ph.GA},
	title = {{The Evolution and Origin of Ionized Gas Velocity Dispersion from z {\ensuremath{\sim}} 2.6 to z {\ensuremath{\sim}} 0.6 with KMOS$^{3D}$}},
	volume = {880},
	year = 2019}

@article{greener20,
	adsurl = {https://ui.adsabs.harvard.edu/abs/2020MNRAS.495.2305G},
	archiveprefix = {arXiv},
	author = {{Greener}, Michael J. and {Arag{\'o}n-Salamanca}, Alfonso and {Merrifield}, Michael R. and {Peterken}, Thomas G. and {Fraser-McKelvie}, Amelia and {Masters}, Karen L. and {Krawczyk}, Coleman M. and {Boardman}, Nicholas F. and {Boquien}, M{\'e}d{\'e}ric and {Andrews}, Brett H. and {Brinkmann}, Jonathan and {Drory}, Niv},
	doi = {10.1093/mnras/staa1300},
	eprint = {2005.02772},
	journal = {\mnras},
	month = jun,
	number = {2},
	pages = {2305-2320},
	primaryclass = {astro-ph.GA},
	title = {{SDSS-IV MaNGA: spatially resolved dust attenuation in spiral galaxies}},
	volume = {495},
	year = 2020}

@article{zibetti22,
	adsurl = {https://ui.adsabs.harvard.edu/abs/2022MNRAS.512.1415Z},
	archiveprefix = {arXiv},
	author = {{Zibetti}, Stefano and {Gallazzi}, Anna R.},
	doi = {10.1093/mnras/stac370},
	eprint = {2202.03975},
	journal = {\mnras},
	month = may,
	number = {1},
	pages = {1415-1429},
	primaryclass = {astro-ph.GA},
	title = {{Stellar mass as the 'glocal' driver of galaxies' stellar population properties}},
	volume = {512},
	year = 2022}

@article{gebek25,
	adsurl = {https://ui.adsabs.harvard.edu/abs/2025A&A...695A..90G},
	archiveprefix = {arXiv},
	author = {{Gebek}, Andrea and {Diemer}, Benedikt and {Martorano}, Marco and {van der Wel}, Arjen and {Pantoni}, Lara and {Baes}, Maarten and {Gabrielpillai}, Austen and {Utsav Kapoor}, Anand and {Osinga}, Calvin and {Nersesian}, Angelos and {Matsumoto}, Kosei and {Gordon}, Karl},
	doi = {10.1051/0004-6361/202452768},
	eid = {A90},
	eprint = {2501.12008},
	journal = {\aap},
	month = mar,
	pages = {A90},
	primaryclass = {astro-ph.GA},
	title = {{The mass-dependent UVJ diagram at cosmic noon: A challenge for galaxy evolution models and dust radiative transfer}},
	volume = {695},
	year = 2025}

@article{shirley19,
	adsurl = {https://ui.adsabs.harvard.edu/abs/2019MNRAS.490..634S},
	archiveprefix = {arXiv},
	author = {{Shirley}, Raphael and {Roehlly}, Yannick and {Hurley}, Peter D. and {Buat}, Veronique and {Campos Varillas}, Mar{\'\i}a del Carmen and {Duivenvoorden}, Steven and {Duncan}, Kenneth J. and {Efstathiou}, Andreas and {Farrah}, Duncan and {Gonz{\'a}lez Solares}, Eduardo and {Malek}, Katarzyna and {Marchetti}, Lucia and {McCheyne}, Ian and {Papadopoulos}, Andreas and {Pons}, Estelle and {Scipioni}, Roberto and {Vaccari}, Mattia and {Oliver}, Seb},
	doi = {10.1093/mnras/stz2509},
	eprint = {1909.04003},
	journal = {\mnras},
	month = nov,
	number = {1},
	pages = {634-656},
	primaryclass = {astro-ph.GA},
	title = {{HELP: a catalogue of 170 million objects, selected at 0.36-4.5 {\ensuremath{\mu}}m, from 1270 deg$^{2}$ of prime extragalactic fields}},
	volume = {490},
	year = 2019}

@inproceedings{wuyts20,
	adsurl = {https://ui.adsabs.harvard.edu/abs/2020IAUS..352..253W},
	author = {{Wuyts}, Stijn and {F{\"o}rster Schreiber}, Natascha M.},
	booktitle = {Uncovering Early Galaxy Evolution in the ALMA and JWST Era},
	doi = {10.1017/S1743921319008421},
	editor = {{da Cunha}, Elisabete and {Hodge}, Jacqueline and {Afonso}, Jos{\'e} and {Pentericci}, Laura and {Sobral}, David},
	month = jan,
	pages = {253-265},
	series = {IAU Symposium},
	title = {{Resolved views on early galaxy evolution}},
	volume = {352},
	year = 2020}

@article{belli19,
	adsurl = {https://ui.adsabs.harvard.edu/abs/2019ApJ...874...17B},
	archiveprefix = {arXiv},
	author = {{Belli}, Sirio and {Newman}, Andrew B. and {Ellis}, Richard S.},
	doi = {10.3847/1538-4357/ab07af},
	eid = {17},
	eprint = {1810.00008},
	journal = {\apj},
	month = mar,
	number = {1},
	pages = {17},
	primaryclass = {astro-ph.GA},
	title = {{MOSFIRE Spectroscopy of Quiescent Galaxies at 1.5 < z < 2.5. II. Star Formation Histories and Galaxy Quenching}},
	volume = {874},
	year = 2019}

@article{jones17,
	adsurl = {https://ui.adsabs.harvard.edu/abs/2017A&A...602A..46J},
	archiveprefix = {arXiv},
	author = {{Jones}, A.~P. and {K{\"o}hler}, M. and {Ysard}, N. and {Bocchio}, M. and {Verstraete}, L.},
	doi = {10.1051/0004-6361/201630225},
	eid = {A46},
	eprint = {1703.00775},
	journal = {\aap},
	month = jun,
	pages = {A46},
	primaryclass = {astro-ph.GA},
	title = {{The global dust modelling framework THEMIS}},
	volume = {602},
	year = 2017}

@article{byrne22,
	adsurl = {https://ui.adsabs.harvard.edu/abs/2022MNRAS.512.5329B},
	archiveprefix = {arXiv},
	author = {{Byrne}, C.~M. and {Stanway}, E.~R. and {Eldridge}, J.~J. and {McSwiney}, L. and {Townsend}, O.~T.},
	doi = {10.1093/mnras/stac807},
	eprint = {2203.13275},
	journal = {\mnras},
	month = jun,
	number = {4},
	pages = {5329-5338},
	primaryclass = {astro-ph.SR},
	title = {{The dependence of theoretical synthetic spectra on {\ensuremath{\alpha}}-enhancement in young, binary stellar populations}},
	volume = {512},
	year = 2022}

@article{dalcanton04,
	adsurl = {https://ui.adsabs.harvard.edu/abs/2004ApJ...608..189D},
	archiveprefix = {arXiv},
	author = {{Dalcanton}, Julianne J. and {Yoachim}, Peter and {Bernstein}, Rebecca A.},
	doi = {10.1086/386358},
	eprint = {astro-ph/0402472},
	journal = {\apj},
	month = jun,
	number = {1},
	pages = {189-207},
	primaryclass = {astro-ph},
	title = {{The Formation of Dust Lanes: Implications for Galaxy Evolution}},
	volume = {608},
	year = 2004}

@article{sanchez-blazquez14,
	adsurl = {https://ui.adsabs.harvard.edu/abs/2014A&A...570A...6S},
	archiveprefix = {arXiv},
	author = {{S{\'a}nchez-Bl{\'a}zquez}, P. and {Rosales-Ortega}, F.~F. and {M{\'e}ndez-Abreu}, J. and {P{\'e}rez}, I. and {S{\'a}nchez}, S.~F. and {Zibetti}, S. and {Aguerri}, J.~A.~L. and {Bland-Hawthorn}, J. and {Catal{\'a}n-Torrecilla}, C. and {Cid Fernandes}, R. and {de Amorim}, A. and {de Lorenzo-Caceres}, A. and {Falc{\'o}n-Barroso}, J. and {Galazzi}, A. and {Garc{\'\i}a Benito}, R. and {Gil de Paz}, A. and {Gonz{\'a}lez Delgado}, R. and {Husemann}, B. and {Iglesias-P{\'a}ramo}, Jorge and {Jungwiert}, B. and {Marino}, R.~A. and {M{\'a}rquez}, I. and {Mast}, D. and {Mendoza}, M.~A. and {Moll{\'a}}, M. and {Papaderos}, P. and {Ruiz-Lara}, T. and {van de Ven}, G. and {Walcher}, C.~J. and {Wisotzki}, L.},
	doi = {10.1051/0004-6361/201423635},
	eid = {A6},
	eprint = {1407.0002},
	journal = {\aap},
	month = oct,
	pages = {A6},
	primaryclass = {astro-ph.GA},
	title = {{Stellar population gradients in galaxy discs from the CALIFA survey. The influence of bars}},
	volume = {570},
	year = 2014}

@article{nersesian25,
	adsurl = {https://ui.adsabs.harvard.edu/abs/2025A&A...695A..86N},
	archiveprefix = {arXiv},
	author = {{Nersesian}, Angelos and {van der Wel}, Arjen and {Gallazzi}, Anna R. and {Kaushal}, Yasha and {Bezanson}, Rachel and {Zibetti}, Stefano and {Bell}, Eric F. and {D'Eugenio}, Francesco and {Leja}, Joel and {Martorano}, Marco and {Wu}, Po-Feng},
	doi = {10.1051/0004-6361/202452662},
	eid = {A86},
	eprint = {2502.03021},
	journal = {\aap},
	month = mar,
	pages = {A86},
	primaryclass = {astro-ph.GA},
	title = {{More is better: Strong constraints on the stellar properties of LEGA-C z {\ensuremath{\sim}} 1 galaxies with Prospector}},
	volume = {695},
	year = 2025}

@article{hodge16,
	adsurl = {https://ui.adsabs.harvard.edu/abs/2016ApJ...833..103H},
	archiveprefix = {arXiv},
	author = {{Hodge}, J.~A. and {Swinbank}, A.~M. and {Simpson}, J.~M. and {Smail}, I. and {Walter}, F. and {Alexander}, D.~M. and {Bertoldi}, F. and {Biggs}, A.~D. and {Brandt}, W.~N. and {Chapman}, S.~C. and {Chen}, C.~C. and {Coppin}, K.~E.~K. and {Cox}, P. and {Dannerbauer}, H. and {Edge}, A.~C. and {Greve}, T.~R. and {Ivison}, R.~J. and {Karim}, A. and {Knudsen}, K.~K. and {Menten}, K.~M. and {Rix}, H. -W. and {Schinnerer}, E. and {Wardlow}, J.~L. and {Weiss}, A. and {van der Werf}, P.},
	doi = {10.3847/1538-4357/833/1/103},
	eid = {103},
	eprint = {1609.09649},
	journal = {\apj},
	month = dec,
	number = {1},
	pages = {103},
	primaryclass = {astro-ph.GA},
	title = {{Kiloparsec-scale Dust Disks in High-redshift Luminous Submillimeter Galaxies}},
	volume = {833},
	year = 2016}

@article{kriek13,
	adsurl = {https://ui.adsabs.harvard.edu/abs/2013ApJ...775L..16K},
	archiveprefix = {arXiv},
	author = {{Kriek}, Mariska and {Conroy}, Charlie},
	doi = {10.1088/2041-8205/775/1/L16},
	eid = {L16},
	eprint = {1308.1099},
	journal = {\apjl},
	month = sep,
	number = {1},
	pages = {L16},
	primaryclass = {astro-ph.CO},
	title = {{The Dust Attenuation Law in Distant Galaxies: Evidence for Variation with Spectral Type}},
	volume = {775},
	year = 2013}

@article{tan24,
	adsurl = {https://ui.adsabs.harvard.edu/abs/2024Natur.636...69T},
	archiveprefix = {arXiv},
	author = {{Tan}, Qing-Hua and {Daddi}, Emanuele and {Magnelli}, Benjamin and {Correa}, Camila A. and {Bournaud}, Fr{\'e}d{\'e}ric and {Adscheid}, Sylvia and {Zhang}, Shao-Bo and {Elbaz}, David and {G{\'o}mez-Guijarro}, Carlos and {Kalita}, Boris S. and {Liu}, Daizhong and {Liu}, Zhaoxuan and {Pety}, J{\'e}r{\^o}me and {Puglisi}, Annagrazia and {Schinnerer}, Eva and {Silverman}, John D. and {Valentino}, Francesco},
	doi = {10.1038/s41586-024-08201-6},
	eprint = {2407.16578},
	journal = {\nat},
	month = dec,
	number = {8041},
	pages = {69-74},
	primaryclass = {astro-ph.GA},
	title = {{In situ spheroid formation in distant submillimetre-bright galaxies}},
	volume = {636},
	year = 2024}

@article{price25,
	adsurl = {https://ui.adsabs.harvard.edu/abs/2025ApJ...980...11P},
	archiveprefix = {arXiv},
	author = {{Price}, Sedona H. and {Suess}, Katherine A. and {Williams}, Christina C. and {Bezanson}, Rachel and {Khullar}, Gourav and {Nelson}, Erica J. and {Wang}, Bingjie and {Weaver}, John R. and {Fujimoto}, Seiji and {Kokorev}, Vasily and {Greene}, Jenny E. and {Brammer}, Gabriel and {Cutler}, Sam E. and {Dayal}, Pratika and {Furtak}, Lukas J. and {Labbe}, Ivo and {Leja}, Joel and {Miller}, Tim B. and {Nanayakkara}, Themiya and {Pan}, Richard and {Whitaker}, Katherine E.},
	doi = {10.3847/1538-4357/ada0b1},
	eid = {11},
	eprint = {2310.02500},
	journal = {\apj},
	month = feb,
	number = {1},
	pages = {11},
	primaryclass = {astro-ph.GA},
	title = {{UNCOVER: The Rest-ultraviolet to Near-infrared Multiwavelength Structures and Dust Distributions of Submillimeter-detected Galaxies in A2744}},
	volume = {980},
	year = 2025}

@article{martorano24,
	adsurl = {https://ui.adsabs.harvard.edu/abs/2024ApJ...972..134M},
	archiveprefix = {arXiv},
	author = {{Martorano}, Marco and {van der Wel}, Arjen and {Baes}, Maarten and {Bell}, Eric F. and {Brammer}, Gabriel and {Franx}, Marijn and {Nersesian}, Angelos},
	doi = {10.3847/1538-4357/ad5c6a},
	eid = {134},
	eprint = {2406.17756},
	journal = {\apj},
	month = sep,
	number = {2},
	pages = {134},
	primaryclass = {astro-ph.GA},
	title = {{The Size{\textendash}Mass Relation at Rest-frame 1.5 {\ensuremath{\mu}}m from JWST/NIRCam in the COSMOS-WEB and PRIMER-COSMOS Fields}},
	volume = {972},
	year = 2024}

@article{martorano25,
	adsurl = {https://ui.adsabs.harvard.edu/abs/2025A&A...694A..76M},
	archiveprefix = {arXiv},
	author = {{Martorano}, M. and {van der Wel}, A. and {Baes}, M. and {Bell}, E.~F. and {Brammer}, G. and {Franx}, M. and {Gebek}, A. and {Meidt}, S.~E. and {Miller}, T.~B. and {Nelson}, E. and {Nersesian}, A. and {Price}, S.~H. and {van Dokkum}, P. and {Whitaker}, K.~E. and {Wuyts}, S.},
	doi = {10.1051/0004-6361/202452919},
	eid = {A76},
	eprint = {2501.02956},
	journal = {\aap},
	month = feb,
	pages = {A76},
	primaryclass = {astro-ph.GA},
	title = {{Evolution of the S{\'e}rsic index up to z = 2.5 from JWST and HST}},
	volume = {694},
	year = 2025}

@article{le-bail24,
	adsurl = {https://ui.adsabs.harvard.edu/abs/2024A&A...688A..53L},
	archiveprefix = {arXiv},
	author = {{Le Bail}, Aur{\'e}lien and {Daddi}, Emanuele and {Elbaz}, David and {Dickinson}, Mark and {Giavalisco}, Mauro and {Magnelli}, Benjamin and {G{\'o}mez-Guijarro}, Carlos and {Kalita}, Boris S. and {Koekemoer}, Anton M. and {Holwerda}, Benne W. and {Bournaud}, Fr{\'e}d{\'e}ric and {de la Vega}, Alexander and {Calabr{\`o}}, Antonello and {Dekel}, Avishai and {Cheng}, Yingjie and {Bisigello}, Laura and {Franco}, Maximilien and {Costantin}, Luca and {Lucas}, Ray A. and {P{\'e}rez-Gonz{\'a}lez}, Pablo G. and {Lu}, Shiying and {Wilkins}, Stephen M. and {Arrabal Haro}, Pablo and {Bagley}, Micaela B. and {Finkelstein}, Steven L. and {Kartaltepe}, Jeyhan S. and {Papovich}, Casey and {Pirzkal}, Nor and {Yung}, L.~Y. Aaron},
	doi = {10.1051/0004-6361/202347465},
	eid = {A53},
	eprint = {2307.07599},
	journal = {\aap},
	month = aug,
	pages = {A53},
	primaryclass = {astro-ph.GA},
	title = {{JWST/CEERS sheds light on dusty star-forming galaxies: Forming bulges, lopsidedness, and outside-in quenching at cosmic noon}},
	volume = {688},
	year = 2024}

@article{lorenz24,
	adsurl = {https://ui.adsabs.harvard.edu/abs/2024ApJ...975..187L},
	archiveprefix = {arXiv},
	author = {{Lorenz}, Brian and {Kriek}, Mariska and {Shapley}, Alice E. and {Sanders}, Ryan L. and {Coil}, Alison L. and {Leja}, Joel and {Mobasher}, Bahram and {Nelson}, Erica and {Price}, Sedona H. and {Reddy}, Naveen A. and {Runco}, Jordan N. and {Suess}, Katherine A. and {Shivaei}, Irene and {Siana}, Brian and {Weisz}, Daniel R.},
	doi = {10.3847/1538-4357/ad7de8},
	eid = {187},
	eprint = {2409.18179},
	journal = {\apj},
	month = nov,
	number = {2},
	pages = {187},
	primaryclass = {astro-ph.GA},
	title = {{Stacking and Analyzing MOSDEF Galaxies by Spectral Types: Implications for Dust Geometry and Galaxy Evolution}},
	volume = {975},
	year = 2024}

@article{cullen18,
	adsurl = {https://ui.adsabs.harvard.edu/abs/2018MNRAS.476.3218C},
	archiveprefix = {arXiv},
	author = {{Cullen}, F. and {McLure}, R.~J. and {Khochfar}, S. and {Dunlop}, J.~S. and {Dalla Vecchia}, C. and {Carnall}, A.~C. and {Bourne}, N. and {Castellano}, M. and {Cimatti}, A. and {Cirasuolo}, M. and {Elbaz}, D. and {Fynbo}, J.~P.~U. and {Garilli}, B. and {Koekemoer}, A. and {Marchi}, F. and {Pentericci}, L. and {Talia}, M. and {Zamorani}, G.},
	doi = {10.1093/mnras/sty469},
	eprint = {1712.01292},
	journal = {\mnras},
	month = may,
	number = {3},
	pages = {3218-3232},
	primaryclass = {astro-ph.GA},
	title = {{The VANDELS survey: dust attenuation in star-forming galaxies at z = 3-4}},
	volume = {476},
	year = 2018}

@article{chapman03,
	adsurl = {https://ui.adsabs.harvard.edu/abs/2003ApJ...599...92C},
	archiveprefix = {arXiv},
	author = {{Chapman}, S.~C. and {Windhorst}, R. and {Odewahn}, S. and {Yan}, H. and {Conselice}, C.},
	doi = {10.1086/379120},
	eprint = {astro-ph/0308197},
	journal = {\apj},
	month = dec,
	number = {1},
	pages = {92-104},
	primaryclass = {astro-ph},
	title = {{Hubble Space Telescope Images of Submillimeter Sources: Large Irregular Galaxies at High Redshift}},
	volume = {599},
	year = 2003}

@article{smail97,
	adsurl = {https://ui.adsabs.harvard.edu/abs/1997ApJ...490L...5S},
	archiveprefix = {arXiv},
	author = {{Smail}, Ian and {Ivison}, R.~J. and {Blain}, A.~W.},
	doi = {10.1086/311017},
	eprint = {astro-ph/9708135},
	journal = {\apjl},
	month = nov,
	number = {1},
	pages = {L5-L8},
	primaryclass = {astro-ph},
	title = {{A Deep Sub-millimeter Survey of Lensing Clusters: A New Window on Galaxy Formation and Evolution}},
	volume = {490},
	year = 1997}

@article{smail02,
	adsurl = {https://ui.adsabs.harvard.edu/abs/2002MNRAS.331..495S},
	archiveprefix = {arXiv},
	author = {{Smail}, Ian and {Ivison}, R.~J. and {Blain}, A.~W. and {Kneib}, J. -P.},
	doi = {10.1046/j.1365-8711.2002.05203.x},
	eprint = {astro-ph/0112100},
	journal = {\mnras},
	month = mar,
	number = {2},
	pages = {495-520},
	primaryclass = {astro-ph},
	title = {{The nature of faint submillimetre-selected galaxies}},
	volume = {331},
	year = 2002}

@article{daddi07,
	adsurl = {https://ui.adsabs.harvard.edu/abs/2007ApJ...670..156D},
	archiveprefix = {arXiv},
	author = {{Daddi}, E. and {Dickinson}, M. and {Morrison}, G. and {Chary}, R. and {Cimatti}, A. and {Elbaz}, D. and {Frayer}, D. and {Renzini}, A. and {Pope}, A. and {Alexander}, D.~M. and {Bauer}, F.~E. and {Giavalisco}, M. and {Huynh}, M. and {Kurk}, J. and {Mignoli}, M.},
	doi = {10.1086/521818},
	eprint = {0705.2831},
	journal = {\apj},
	month = nov,
	number = {1},
	pages = {156-172},
	primaryclass = {astro-ph},
	title = {{Multiwavelength Study of Massive Galaxies at z\raisebox{-0.5ex}\textasciitilde2. I. Star Formation and Galaxy Growth}},
	volume = {670},
	year = 2007}

@article{akins22,
	adsurl = {https://ui.adsabs.harvard.edu/abs/2022ApJ...929...94A},
	archiveprefix = {arXiv},
	author = {{Akins}, Hollis B. and {Narayanan}, Desika and {Whitaker}, Katherine E. and {Dav{\'e}}, Romeel and {Lower}, Sidney and {Bezanson}, Rachel and {Feldmann}, Robert and {Kriek}, Mariska},
	doi = {10.3847/1538-4357/ac5d3a},
	eid = {94},
	eprint = {2105.12748},
	journal = {\apj},
	month = apr,
	number = {1},
	pages = {94},
	primaryclass = {astro-ph.GA},
	title = {{Quenching and the UVJ Diagram in the SIMBA Cosmological Simulation}},
	volume = {929},
	year = 2022}

@article{donnari19,
	adsurl = {https://ui.adsabs.harvard.edu/abs/2019MNRAS.485.4817D},
	archiveprefix = {arXiv},
	author = {{Donnari}, Martina and {Pillepich}, Annalisa and {Nelson}, Dylan and {Vogelsberger}, Mark and {Genel}, Shy and {Weinberger}, Rainer and {Marinacci}, Federico and {Springel}, Volker and {Hernquist}, Lars},
	doi = {10.1093/mnras/stz712},
	eprint = {1812.07584},
	journal = {\mnras},
	month = jun,
	number = {4},
	pages = {4817-4840},
	primaryclass = {astro-ph.GA},
	title = {{The star formation activity of IllustrisTNG galaxies: main sequence, UVJ diagram, quenched fractions, and systematics}},
	volume = {485},
	year = 2019}

@article{wang17,
	adsurl = {https://ui.adsabs.harvard.edu/abs/2017MNRAS.469.4063W},
	archiveprefix = {arXiv},
	author = {{Wang}, Weichen and {Faber}, S.~M. and {Liu}, F.~S. and {Guo}, Yicheng and {Pacifici}, Camilla and {Koo}, David C. and {Kassin}, Susan A. and {Mao}, Shude and {Fang}, Jerome J. and {Chen}, Zhu and {Koekemoer}, Anton M. and {Kocevski}, Dale D. and {Ashby}, M.~L.~N.},
	doi = {10.1093/mnras/stx1148},
	eprint = {1705.05404},
	journal = {\mnras},
	month = aug,
	number = {4},
	pages = {4063-4082},
	primaryclass = {astro-ph.GA},
	title = {{UVI colour gradients of 0.4 < z < 1.4 star-forming main-sequence galaxies in CANDELS: dust extinction and star formation profiles}},
	volume = {469},
	year = 2017}

@article{lin24,
	adsurl = {https://ui.adsabs.harvard.edu/abs/2024ApJ...977..175L},
	archiveprefix = {arXiv},
	author = {{Lin}, Lin and {Shen}, Shiyin and {Yesuf}, Hassen M. and {Mao}, Ye-Wei and {Hao}, Lei},
	doi = {10.3847/1538-4357/ad8a61},
	eid = {175},
	eprint = {2410.16651},
	journal = {\apj},
	month = dec,
	number = {2},
	pages = {175},
	primaryclass = {astro-ph.GA},
	title = {{Radial Profiles of {\ensuremath{\Sigma}}$_{*}$, {\ensuremath{\Sigma}}$_{SFR}$, Gas Metallicity, and Their Correlations across the Galactic Mass{\textendash}Size Plane}},
	volume = {977},
	year = 2024}

@article{goddard17,
	adsurl = {https://ui.adsabs.harvard.edu/abs/2017MNRAS.466.4731G},
	archiveprefix = {arXiv},
	author = {{Goddard}, D. and {Thomas}, D. and {Maraston}, C. and {Westfall}, K. and {Etherington}, J. and {Riffel}, R. and {Mallmann}, N.~D. and {Zheng}, Z. and {Argudo-Fern{\'a}ndez}, M. and {Lian}, J. and {Bershady}, M. and {Bundy}, K. and {Drory}, N. and {Law}, D. and {Yan}, R. and {Wake}, D. and {Weijmans}, A. and {Bizyaev}, D. and {Brownstein}, J. and {Lane}, R.~R. and {Maiolino}, R. and {Masters}, K. and {Merrifield}, M. and {Nitschelm}, C. and {Pan}, K. and {Roman-Lopes}, A. and {Storchi-Bergmann}, T. and {Schneider}, D.~P.},
	doi = {10.1093/mnras/stw3371},
	eprint = {1612.01546},
	journal = {\mnras},
	month = apr,
	number = {4},
	pages = {4731-4758},
	primaryclass = {astro-ph.GA},
	title = {{SDSS-IV MaNGA: Spatially resolved star formation histories in galaxies as a function of galaxy mass and type}},
	volume = {466},
	year = 2017}

@article{belfiore17,
	adsurl = {https://ui.adsabs.harvard.edu/abs/2017MNRAS.466.2570B},
	archiveprefix = {arXiv},
	author = {{Belfiore}, Francesco and {Maiolino}, Roberto and {Maraston}, Claudia and {Emsellem}, Eric and {Bershady}, Matthew A. and {Masters}, Karen L. and {Bizyaev}, Dmitry and {Boquien}, M{\'e}d{\'e}ric and {Brownstein}, Joel R. and {Bundy}, Kevin and {Diamond-Stanic}, Aleksandar M. and {Drory}, Niv and {Heckman}, Timothy M. and {Law}, David R. and {Malanushenko}, Olena and {Oravetz}, Audrey and {Pan}, Kaike and {Roman-Lopes}, Alexandre and {Thomas}, Daniel and {Weijmans}, Anne-Marie and {Westfall}, Kyle B. and {Yan}, Renbin},
	doi = {10.1093/mnras/stw3211},
	eprint = {1609.01737},
	journal = {\mnras},
	month = apr,
	number = {3},
	pages = {2570-2589},
	primaryclass = {astro-ph.GA},
	title = {{SDSS-IV MaNGA - the spatially resolved transition from star formation to quiescence}},
	volume = {466},
	year = 2017}

@article{lin17,
	adsurl = {https://ui.adsabs.harvard.edu/abs/2017ApJ...851...18L},
	archiveprefix = {arXiv},
	author = {{Lin}, Lihwai and {Belfiore}, Francesco and {Pan}, Hsi-An and {Bothwell}, M.~S. and {Hsieh}, Pei-Ying and {Huang}, Shan and {Xiao}, Ting and {S{\'a}nchez}, Sebasti{\'a}n F. and {Hsieh}, Bau-Ching and {Masters}, Karen and {Ramya}, S. and {Lin}, Jing-Hua and {Hsu}, Chin-Hao and {Li}, Cheng and {Maiolino}, Roberto and {Bundy}, Kevin and {Bizyaev}, Dmitry and {Drory}, Niv and {Ibarra-Medel}, H{\'e}ctor and {Lacerna}, Ivan and {Haines}, Tim and {Smethurst}, Rebecca and {Stark}, David V. and {Thomas}, Daniel},
	doi = {10.3847/1538-4357/aa96ae},
	eid = {18},
	eprint = {1710.08610},
	journal = {\apj},
	month = dec,
	number = {1},
	pages = {18},
	primaryclass = {astro-ph.GA},
	title = {{SDSS-IV MaNGA-resolved Star Formation and Molecular Gas Properties of Green Valley Galaxies: A First Look with ALMA and MaNGA}},
	volume = {851},
	year = 2017}

@article{lin19,
	adsurl = {https://ui.adsabs.harvard.edu/abs/2019ApJ...872...50L},
	archiveprefix = {arXiv},
	author = {{Lin}, Lihwai and {Hsieh}, Bau-Ching and {Pan}, Hsi-An and {Rembold}, Sandro B. and {S{\'a}nchez}, Sebasti{\'a}n F. and {Argudo-Fern{\'a}ndez}, Maria and {Rowlands}, Kate and {Belfiore}, Francesco and {Bizyaev}, Dmitry and {Lacerna}, Ivan and {Riffel}, Rogr{\'e}io and {Rong}, Yu and {Yuan}, Fangting and {Drory}, Niv and {Maiolino}, Roberto and {Wilcots}, Eric},
	doi = {10.3847/1538-4357/aafa84},
	eid = {50},
	eprint = {1901.05126},
	journal = {\apj},
	month = feb,
	number = {1},
	pages = {50},
	primaryclass = {astro-ph.GA},
	title = {{SDSS-IV MaNGA: Inside-out versus Outside-in Quenching of Galaxies in Different Local Environments}},
	volume = {872},
	year = 2019}

@article{ellison18,
	adsurl = {https://ui.adsabs.harvard.edu/abs/2018MNRAS.474.2039E},
	archiveprefix = {arXiv},
	author = {{Ellison}, Sara L. and {S{\'a}nchez}, Sebastian F. and {Ibarra-Medel}, Hector and {Antonio}, Braulio and {Mendel}, J. Trevor and {Barrera-Ballesteros}, Jorge},
	doi = {10.1093/mnras/stx2882},
	eprint = {1711.00915},
	journal = {\mnras},
	month = feb,
	number = {2},
	pages = {2039-2054},
	primaryclass = {astro-ph.GA},
	title = {{Star formation is boosted (and quenched) from the inside-out: radial star formation profiles from MaNGA}},
	volume = {474},
	year = 2018}

@article{belfiore18,
	adsurl = {https://ui.adsabs.harvard.edu/abs/2018MNRAS.477.3014B},
	archiveprefix = {arXiv},
	author = {{Belfiore}, Francesco and {Maiolino}, Roberto and {Bundy}, Kevin and {Masters}, Karen and {Bershady}, Matthew and {Oyarz{\'u}n}, Grecco A. and {Lin}, Lihwai and {Cano-Diaz}, Mariana and {Wake}, David and {Spindler}, Ashley and {Thomas}, Daniel and {Brownstein}, Joel R. and {Drory}, Niv and {Yan}, Renbin},
	doi = {10.1093/mnras/sty768},
	eprint = {1710.05034},
	journal = {\mnras},
	month = jul,
	number = {3},
	pages = {3014-3029},
	primaryclass = {astro-ph.GA},
	title = {{SDSS IV MaNGA - sSFR profiles and the slow quenching of discs in green valley galaxies}},
	volume = {477},
	year = 2018}

@article{liu17,
	adsurl = {https://ui.adsabs.harvard.edu/abs/2017ApJ...844L...2L},
	archiveprefix = {arXiv},
	author = {{Liu}, F.~S. and {Jiang}, Dongfei and {Faber}, S.~M. and {Koo}, David C. and {Yesuf}, Hassen M. and {Tacchella}, Sandro and {Mao}, Shude and {Wang}, Weichen and {Guo}, Yicheng and {Fang}, Jerome J. and {Barro}, Guillermo and {Zheng}, Xianzhong and {Jia}, Meng and {Tong}, Wei and {Liu}, Lu and {Meng}, Xianmin},
	doi = {10.3847/2041-8213/aa7cf5},
	eid = {L2},
	eprint = {1707.00226},
	journal = {\apjl},
	month = jul,
	number = {1},
	pages = {L2},
	primaryclass = {astro-ph.GA},
	title = {{The Origins of UV-optical Color Gradients in Star-forming Galaxies at z {\ensuremath{\sim}} 2: Predominant Dust Gradients but Negligible sSFR Gradients}},
	volume = {844},
	year = 2017}

@article{liu16,
	adsurl = {https://ui.adsabs.harvard.edu/abs/2016ApJ...822L..25L},
	archiveprefix = {arXiv},
	author = {{Liu}, F.~S. and {Jiang}, Dongfei and {Guo}, Yicheng and {Koo}, David C. and {Faber}, S.~M. and {Zheng}, Xianzhong and {Yesuf}, Hassen M. and {Barro}, Guillermo and {Li}, Yao and {Li}, Dingpeng and {Wang}, Weichen and {Mao}, Shude and {Fang}, Jerome J.},
	doi = {10.3847/2041-8205/822/2/L25},
	eid = {L25},
	eprint = {1604.05780},
	journal = {\apjl},
	month = may,
	number = {2},
	pages = {L25},
	primaryclass = {astro-ph.GA},
	title = {{The UV-Optical Color Gradients in Star-forming Galaxies at 0.5 < z < 1.5: Origins and Link to Galaxy Assembly}},
	volume = {822},
	year = 2016}

@article{wu05,
	adsurl = {https://ui.adsabs.harvard.edu/abs/2005ApJ...622..244W},
	archiveprefix = {arXiv},
	author = {{Wu}, Hong and {Shao}, Zhengyi and {Mo}, H.~J. and {Xia}, Xiaoyang and {Deng}, Zugan},
	doi = {10.1086/427821},
	eprint = {astro-ph/0404226},
	journal = {\apj},
	month = mar,
	number = {1},
	pages = {244-259},
	primaryclass = {astro-ph},
	title = {{Optical and Near-Infrared Color Profiles in Nearby Early-Type Galaxies and the Implied Age and Metallicity Gradients}},
	volume = {622},
	year = 2005}

@article{tortora11,
	adsurl = {https://ui.adsabs.harvard.edu/abs/2011MNRAS.418.1557T},
	archiveprefix = {arXiv},
	author = {{Tortora}, C. and {Napolitano}, N.~R. and {Romanowsky}, A.~J. and {Jetzer}, Ph. and {Cardone}, V.~F. and {Capaccioli}, M.},
	doi = {10.1111/j.1365-2966.2011.19438.x},
	eprint = {1107.2918},
	journal = {\mnras},
	month = dec,
	number = {3},
	pages = {1557-1564},
	primaryclass = {astro-ph.CO},
	title = {{Stellar mass-to-light ratio gradients in galaxies: correlations with mass}},
	volume = {418},
	year = 2011}

@article{gillman24,
	adsurl = {https://ui.adsabs.harvard.edu/abs/2024A&A...691A.299G},
	archiveprefix = {arXiv},
	author = {{Gillman}, Steven and {Smail}, Ian and {Gullberg}, Bitten and {Swinbank}, A.~M. and {Vijayan}, Aswin P. and {Lee}, Minju and {Brammer}, Gabe and {Dudzevi{\v{c}}i{\={u}}t{\.{e}}}, Ugn{\.{e}} and {Greve}, Thomas R. and {Almaini}, Omar and {Brinch}, Malte and {Chapman}, Scott C. and {Chen}, Chian-Chou and {Ikarashi}, Soh and {Matsuda}, Yuichi and {Wang}, Wei-Hao and {Walter}, Fabian and {van der Werf}, Paul P.},
	doi = {10.1051/0004-6361/202451006},
	eid = {A299},
	eprint = {2406.03544},
	journal = {\aap},
	month = nov,
	pages = {A299},
	primaryclass = {astro-ph.GA},
	title = {{The structure of massive star-forming galaxies from JWST and ALMA: Dusty, high-redshift disc galaxies}},
	volume = {691},
	year = 2024}

@article{adscheid24,
	adsurl = {https://ui.adsabs.harvard.edu/abs/2024A&A...685A...1A},
	archiveprefix = {arXiv},
	author = {{Adscheid}, Sylvia and {Magnelli}, Benjamin and {Liu}, Daizhong and {Bertoldi}, Frank and {Delvecchio}, Ivan and {Gruppioni}, Carlotta and {Schinnerer}, Eva and {Traina}, Alberto and {B{\'e}thermin}, Matthieu and {Gkogkou}, Athanasia},
	doi = {10.1051/0004-6361/202348407},
	eid = {A1},
	eprint = {2403.03125},
	journal = {\aap},
	month = may,
	pages = {A1},
	primaryclass = {astro-ph.GA},
	title = {{A$^{3}$COSMOS and A$^{3}$GOODSS: Continuum source catalogues and multi-band number counts}},
	volume = {685},
	year = 2024}

@article{williams23,
	adsurl = {https://ui.adsabs.harvard.edu/abs/2023ApJS..268...64W},
	archiveprefix = {arXiv},
	author = {{Williams}, Christina C. and {Tacchella}, Sandro and {Maseda}, Michael V. and {Robertson}, Brant E. and {Johnson}, Benjamin D. and {Willott}, Chris J. and {Eisenstein}, Daniel J. and {Willmer}, Christopher N.~A. and {Ji}, Zhiyuan and {Hainline}, Kevin N. and {Helton}, Jakob M. and {Alberts}, Stacey and {Baum}, Stefi and {Bhatawdekar}, Rachana and {Boyett}, Kristan and {Bunker}, Andrew J. and {Carniani}, Stefano and {Charlot}, Stephane and {Chevallard}, Jacopo and {Curtis-Lake}, Emma and {de Graaff}, Anna and {Egami}, Eiichi and {Franx}, Marijn and {Kumari}, Nimisha and {Maiolino}, Roberto and {Nelson}, Erica J. and {Rieke}, Marcia J. and {Sandles}, Lester and {Shivaei}, Irene and {Simmonds}, Charlotte and {Smit}, Renske and {Suess}, Katherine A. and {Sun}, Fengwu and {{\"U}bler}, Hannah and {Witstok}, Joris},
	doi = {10.3847/1538-4365/acf130},
	eid = {64},
	eprint = {2301.09780},
	journal = {\apjs},
	month = oct,
	number = {2},
	pages = {64},
	primaryclass = {astro-ph.GA},
	title = {{JEMS: A Deep Medium-band Imaging Survey in the Hubble Ultra Deep Field with JWST NIRCam and NIRISS}},
	volume = {268},
	year = 2023}

@article{oesch23,
	adsurl = {https://ui.adsabs.harvard.edu/abs/2023MNRAS.525.2864O},
	archiveprefix = {arXiv},
	author = {{Oesch}, P.~A. and {Brammer}, G. and {Naidu}, R.~P. and {Bouwens}, R.~J. and {Chisholm}, J. and {Illingworth}, G.~D. and {Matthee}, J. and {Nelson}, E. and {Qin}, Y. and {Reddy}, N. and {Shapley}, A. and {Shivaei}, I. and {van Dokkum}, P. and {Weibel}, A. and {Whitaker}, K. and {Wuyts}, S. and {Covelo-Paz}, A. and {Endsley}, R. and {Fudamoto}, Y. and {Giovinazzo}, E. and {Herard-Demanche}, T. and {Kerutt}, J. and {Kramarenko}, I. and {Labbe}, I. and {Leonova}, E. and {Lin}, J. and {Magee}, D. and {Marchesini}, D. and {Maseda}, M. and {Mason}, C. and {Matharu}, J. and {Meyer}, R.~A. and {Neufeld}, C. and {Prieto Lyon}, G. and {Schaerer}, D. and {Sharma}, R. and {Shuntov}, M. and {Smit}, R. and {Stefanon}, M. and {Wyithe}, J.~S.~B. and {Xiao}, M.},
	doi = {10.1093/mnras/stad2411},
	eprint = {2304.02026},
	journal = {\mnras},
	month = oct,
	number = {2},
	pages = {2864-2874},
	primaryclass = {astro-ph.GA},
	title = {{The JWST FRESCO survey: legacy NIRCam/grism spectroscopy and imaging in the two GOODS fields}},
	volume = {525},
	year = 2023}

@article{nedkova24a,
	adsurl = {https://ui.adsabs.harvard.edu/abs/2024ApJ...970..188N},
	archiveprefix = {arXiv},
	author = {{Nedkova}, Kalina V. and {Rafelski}, Marc and {Teplitz}, Harry I. and {Mehta}, Vihang and {Degroot}, Laura and {Ravindranath}, Swara and {Alavi}, Anahita and {Beckett}, Alexander and {Grogin}, Norman A. and {H{\"a}u{\ss}ler}, Boris and {Koekemoer}, Anton M. and {Oyarz{\'u}n}, Grecco A. and {Prichard}, Laura and {Revalski}, Mitchell and {Snyder}, Gregory F. and {Sunnquist}, Ben and {Wang}, Xin and {Windhorst}, Rogier A. and {Chartab}, Nima and {Conselice}, Christopher J. and {Guo}, Yicheng and {Hathi}, Nimish and {Hayes}, Matthew J. and {Ji}, Zhiyuan and {Kim}, Keunho J. and {Lucas}, Ray A. and {Mobasher}, Bahram and {O'Connell}, Robert W. and {Sattari}, Zahra and {Smith}, Brent M. and {Taamoli}, Sina and {Yung}, L.~Y. Aaron and {The Uvcandels Team}},
	doi = {10.3847/1538-4357/ad4ede},
	eid = {188},
	eprint = {2405.10908},
	journal = {\apj},
	month = aug,
	number = {2},
	pages = {188},
	primaryclass = {astro-ph.GA},
	title = {{UVCANDELS: The Role of Dust on the Stellar Mass{\textendash}Size Relation of Disk Galaxies at 0.5 {\ensuremath{\leq}} z {\ensuremath{\leq}} 3.0}},
	volume = {970},
	year = 2024}

@article{conroy10,
	adsurl = {https://ui.adsabs.harvard.edu/abs/2010ApJ...712..833C},
	archiveprefix = {arXiv},
	author = {{Conroy}, Charlie and {Gunn}, James E.},
	doi = {10.1088/0004-637X/712/2/833},
	eprint = {0911.3151},
	journal = {\apj},
	month = apr,
	number = {2},
	pages = {833-857},
	primaryclass = {astro-ph.CO},
	title = {{The Propagation of Uncertainties in Stellar Population Synthesis Modeling. III. Model Calibration, Comparison, and Evaluation}},
	volume = {712},
	year = 2010}

@article{carnall23,
	adsurl = {https://ui.adsabs.harvard.edu/abs/2023Natur.619..716C},
	archiveprefix = {arXiv},
	author = {{Carnall}, Adam C. and {McLure}, Ross J. and {Dunlop}, James S. and {McLeod}, Derek J. and {Wild}, Vivienne and {Cullen}, Fergus and {Magee}, Dan and {Begley}, Ryan and {Cimatti}, Andrea and {Donnan}, Callum T. and {Hamadouche}, Massissilia L. and {Jewell}, Sophie M. and {Walker}, Sam},
	doi = {10.1038/s41586-023-06158-6},
	eprint = {2301.11413},
	journal = {\nat},
	month = jul,
	number = {7971},
	pages = {716-719},
	primaryclass = {astro-ph.GA},
	title = {{A massive quiescent galaxy at redshift 4.658}},
	volume = {619},
	year = 2023}

@article{killi24,
	adsurl = {https://ui.adsabs.harvard.edu/abs/2024A&A...691A..52K},
	archiveprefix = {arXiv},
	author = {{Killi}, Meghana and {Watson}, Darach and {Brammer}, Gabriel and {McPartland}, Conor and {Antwi-Danso}, Jacqueline and {Newshore}, Rosa and {Coe}, Dan and {Allen}, Natalie and {Fynbo}, Johan P.~U. and {Gould}, Katriona and {Heintz}, Kasper E. and {Rusakov}, Vadim and {Vejlgaard}, Simone},
	doi = {10.1051/0004-6361/202348857},
	eid = {A52},
	eprint = {2312.03065},
	journal = {\aap},
	month = nov,
	pages = {A52},
	primaryclass = {astro-ph.GA},
	title = {{Deciphering the JWST spectrum of a 'little red dot' at z {\ensuremath{\sim}} 4.53: An obscured AGN and its star-forming host}},
	volume = {691},
	year = 2024}

@article{oke83,
	adsurl = {https://ui.adsabs.harvard.edu/abs/1983ApJ...266..713O},
	author = {{Oke}, J.~B. and {Gunn}, J.~E.},
	doi = {10.1086/160817},
	journal = {\apj},
	month = mar,
	pages = {713-717},
	title = {{Secondary standard stars for absolute spectrophotometry.}},
	volume = {266},
	year = 1983}

@article{suess20,
	adsurl = {https://ui.adsabs.harvard.edu/abs/2020ApJ...899L..26S},
	archiveprefix = {arXiv},
	author = {{Suess}, Katherine A. and {Kriek}, Mariska and {Price}, Sedona H. and {Barro}, Guillermo},
	doi = {10.3847/2041-8213/abacc9},
	eid = {L26},
	eprint = {2008.02817},
	journal = {\apjl},
	month = aug,
	number = {2},
	pages = {L26},
	primaryclass = {astro-ph.GA},
	title = {{Color Gradients along the Quiescent Galaxy Sequence: Clues to Quenching and Structural Growth}},
	volume = {899},
	year = 2020}

@article{cutler24,
	adsurl = {https://ui.adsabs.harvard.edu/abs/2024ApJ...967L..23C},
	archiveprefix = {arXiv},
	author = {{Cutler}, Sam E. and {Whitaker}, Katherine E. and {Weaver}, John R. and {Wang}, Bingjie and {Pan}, Richard and {Bezanson}, Rachel and {Furtak}, Lukas J. and {Labbe}, Ivo and {Leja}, Joel and {Price}, Sedona H. and {Cheng}, Yingjie and {Clausen}, Maike and {Cullen}, Fergus and {Dayal}, Pratika and {de Graaff}, Anna and {Dickinson}, Mark and {Dunlop}, James S. and {Feldmann}, Robert and {Franx}, Marijn and {Giavalisco}, Mauro and {Glazebrook}, Karl and {Greene}, Jenny E. and {Grogin}, Norman A. and {Illingworth}, Garth and {Koekemoer}, Anton M. and {Kokorev}, Vasily and {Marchesini}, Danilo and {Maseda}, Michael V. and {Miller}, Tim B. and {Nanayakkara}, Themiya and {Nelson}, Erica J. and {Setton}, David J. and {Shipley}, Heath and {Suess}, Katherine A.},
	doi = {10.3847/2041-8213/ad464c},
	eid = {L23},
	eprint = {2312.15012},
	journal = {\apjl},
	month = jun,
	number = {2},
	pages = {L23},
	primaryclass = {astro-ph.GA},
	title = {{Two Distinct Classes of Quiescent Galaxies at Cosmic Noon Revealed by JWST PRIMER and UNCOVER}},
	volume = {967},
	year = 2024}

@article{ormerod24,
	adsurl = {https://ui.adsabs.harvard.edu/abs/2024MNRAS.527.6110O},
	archiveprefix = {arXiv},
	author = {{Ormerod}, K. and {Conselice}, C.~J. and {Adams}, N.~J. and {Harvey}, T. and {Austin}, D. and {Trussler}, J. and {Ferreira}, L. and {Caruana}, J. and {Lucatelli}, G. and {Li}, Q. and {Roper}, W.~J.},
	doi = {10.1093/mnras/stad3597},
	eprint = {2309.04377},
	journal = {\mnras},
	month = jan,
	number = {3},
	pages = {6110-6125},
	primaryclass = {astro-ph.GA},
	title = {{EPOCHS VI: the size and shape evolution of galaxies since z 8 with JWST Observations}},
	volume = {527},
	year = 2024}

@article{ito24,
	adsurl = {https://ui.adsabs.harvard.edu/abs/2024ApJ...964..192I},
	archiveprefix = {arXiv},
	author = {{Ito}, Kei and {Valentino}, Francesco and {Brammer}, Gabriel and {Faisst}, Andreas L. and {Gillman}, Steven and {G{\'o}mez-Guijarro}, Carlos and {Gould}, Katriona M.~L. and {Heintz}, Kasper E. and {Ilbert}, Olivier and {Jespersen}, Christian Kragh and {Kokorev}, Vasily and {Kubo}, Mariko and {Magdis}, Georgios E. and {McPartland}, Conor J.~R. and {Onodera}, Masato and {Rizzo}, Francesca and {Tanaka}, Masayuki and {Toft}, Sune and {Vijayan}, Aswin P. and {Weaver}, John R. and {Whitaker}, Katherine E. and {Wright}, Lillian},
	doi = {10.3847/1538-4357/ad2512},
	eid = {192},
	eprint = {2307.06994},
	journal = {\apj},
	month = apr,
	number = {2},
	pages = {192},
	primaryclass = {astro-ph.GA},
	title = {{Size{\textendash}Stellar Mass Relation and Morphology of Quiescent Galaxies at z {\ensuremath{\geq}} 3 in Public JWST Fields}},
	volume = {964},
	year = 2024}

@article{miller22,
	adsurl = {https://ui.adsabs.harvard.edu/abs/2022ApJ...941L..37M},
	archiveprefix = {arXiv},
	author = {{Miller}, Tim B. and {Whitaker}, Katherine E. and {Nelson}, Erica J. and {van Dokkum}, Pieter and {Bezanson}, Rachel and {Brammer}, Gabriel and {Heintz}, Kasper E. and {Leja}, Joel and {Suess}, Katherine A. and {Weaver}, John R.},
	doi = {10.3847/2041-8213/aca675},
	eid = {L37},
	eprint = {2209.12954},
	journal = {\apjl},
	month = dec,
	number = {2},
	pages = {L37},
	primaryclass = {astro-ph.GA},
	title = {{Early JWST Imaging Reveals Strong Optical and NIR Color Gradients in Galaxies at z 2 Driven Mostly by Dust}},
	volume = {941},
	year = 2022}

@article{van-der-wel24,
	adsurl = {https://ui.adsabs.harvard.edu/abs/2024ApJ...960...53V},
	archiveprefix = {arXiv},
	author = {{van der Wel}, Arjen and {Martorano}, Marco and {H{\"a}u{\ss}ler}, Boris and {Nedkova}, Kalina V. and {Miller}, Tim B. and {Brammer}, Gabriel B. and {van de Ven}, Glenn and {Leja}, Joel and {Bezanson}, Rachel S. and {Muzzin}, Adam and {Marchesini}, Danilo and {de Graaff}, Anna and {Nelson}, Erica J. and {Kriek}, Mariska and {Bell}, Eric F. and {Franx}, Marijn},
	doi = {10.3847/1538-4357/ad02ee},
	eid = {53},
	eprint = {2307.03264},
	journal = {\apj},
	month = jan,
	number = {1},
	pages = {53},
	primaryclass = {astro-ph.GA},
	title = {{Stellar Half-mass Radii of 0.5 z < 2.3 Galaxies: Comparison with JWST/NIRCam Half-light Radii}},
	volume = {960},
	year = 2024}

@article{martorano23,
	adsurl = {https://ui.adsabs.harvard.edu/abs/2023ApJ...957...46M},
	archiveprefix = {arXiv},
	author = {{Martorano}, Marco and {van der Wel}, Arjen and {Bell}, Eric F. and {Franx}, Marijn and {Whitaker}, Katherine E. and {Nersesian}, Angelos and {Price}, Sedona H. and {Baes}, Maarten and {Suess}, Katherine A. and {Nelson}, Erica J. and {Miller}, Tim B. and {Bezanson}, Rachel and {Brammer}, Gabriel},
	doi = {10.3847/1538-4357/acf716},
	eid = {46},
	eprint = {2308.11392},
	journal = {\apj},
	month = nov,
	number = {1},
	pages = {46},
	primaryclass = {astro-ph.GA},
	title = {{Rest-frame Near-infrared Radial Light Profiles up to z = 3 from JWST/NIRCam: Wavelength Dependence of the S{\'e}rsic Index}},
	volume = {957},
	year = 2023}

@article{eisenstein23,
	adsurl = {https://ui.adsabs.harvard.edu/abs/2023arXiv230602465E},
	archiveprefix = {arXiv},
	author = {{Eisenstein}, Daniel J. and {Willott}, Chris and {Alberts}, Stacey and {Arribas}, Santiago and {Bonaventura}, Nina and {Bunker}, Andrew J. and {Cameron}, Alex J. and {Carniani}, Stefano and {Charlot}, Stephane and {Curtis-Lake}, Emma and {D'Eugenio}, Francesco and {Endsley}, Ryan and {Ferruit}, Pierre and {Giardino}, Giovanna and {Hainline}, Kevin and {Hausen}, Ryan and {Jakobsen}, Peter and {Johnson}, Benjamin D. and {Maiolino}, Roberto and {Rieke}, Marcia and {Rieke}, George and {Rix}, Hans-Walter and {Robertson}, Brant and {Stark}, Daniel P. and {Tacchella}, Sandro and {Williams}, Christina C. and {Willmer}, Christopher N.~A. and {Baker}, William M. and {Baum}, Stefi and {Bhatawdekar}, Rachana and {Boyett}, Kristan and {Chen}, Zuyi and {Chevallard}, Jacopo and {Circosta}, Chiara and {Curti}, Mirko and {Danhaive}, A. Lola and {DeCoursey}, Christa and {de Graaff}, Anna and {Dressler}, Alan and {Egami}, Eiichi and {Helton}, Jakob M. and {Hviding}, Raphael E. and {Ji}, Zhiyuan and {Jones}, Gareth C. and {Kumari}, Nimisha and {L{\"u}tzgendorf}, Nora and {Laseter}, Isaac and {Looser}, Tobias J. and {Lyu}, Jianwei and {Maseda}, Michael V. and {Nelson}, Erica and {Parlanti}, Eleonora and {Perna}, Michele and {Pusk{\'a}s}, D{\'a}vid and {Rawle}, Tim and {Rodr{\'\i}guez Del Pino}, Bruno and {Sandles}, Lester and {Saxena}, Aayush and {Scholtz}, Jan and {Sharpe}, Katherine and {Shivaei}, Irene and {Silcock}, Maddie S. and {Simmonds}, Charlotte and {Skarbinski}, Maya and {Smit}, Renske and {Stone}, Meredith and {Suess}, Katherine A. and {Sun}, Fengwu and {Tang}, Mengtao and {Topping}, Michael W. and {{\"U}bler}, Hannah and {Villanueva}, Natalia C. and {Wallace}, Imaan E.~B. and {Whitler}, Lily and {Witstok}, Joris and {Woodrum}, Charity},
	doi = {10.48550/arXiv.2306.02465},
	eid = {arXiv:2306.02465},
	eprint = {2306.02465},
	journal = {arXiv e-prints},
	month = jun,
	pages = {arXiv:2306.02465},
	primaryclass = {astro-ph.GA},
	title = {{Overview of the JWST Advanced Deep Extragalactic Survey (JADES)}},
	year = 2023}

@article{nelson19,
	adsurl = {https://ui.adsabs.harvard.edu/abs/2019ApJ...870..130N},
	archiveprefix = {arXiv},
	author = {{Nelson}, Erica J. and {Tadaki}, Ken-ichi and {Tacconi}, Linda J. and {Lutz}, Dieter and {F{\"o}rster Schreiber}, Natascha M. and {Cibinel}, Anna and {Wuyts}, Stijn and {Lang}, Philipp and {Leja}, Joel and {Montes}, Mireia and {Oesch}, Pascal A. and {Belli}, Sirio and {Davies}, Rebecca L. and {Davies}, Richard I. and {Genzel}, Reinhard and {Lippa}, Magdalena and {Price}, Sedona H. and {{\"U}bler}, Hannah and {Wisnioski}, Emily},
	doi = {10.3847/1538-4357/aaf38a},
	eid = {130},
	eprint = {1801.02647},
	journal = {\apj},
	month = jan,
	number = {2},
	pages = {130},
	primaryclass = {astro-ph.GA},
	title = {{Millimeter Mapping at z {\ensuremath{\sim}} 1: Dust-obscured Bulge Building and Disk Growth}},
	volume = {870},
	year = 2019}

@article{mosleh20,
	adsurl = {https://ui.adsabs.harvard.edu/abs/2020ApJ...905..170M},
	archiveprefix = {arXiv},
	author = {{Mosleh}, Moein and {Hosseinnejad}, Shiva and {Hosseini-ShahiSavandi}, S. Zahra and {Tacchella}, Sandro},
	doi = {10.3847/1538-4357/abc7cc},
	eid = {170},
	eprint = {2011.04656},
	journal = {\apj},
	month = dec,
	number = {2},
	pages = {170},
	primaryclass = {astro-ph.GA},
	title = {{Galaxy Sizes Since z = 2 from the Perspective of Stellar Mass Distribution within Galaxies}},
	volume = {905},
	year = 2020}

@article{peletier90,
	adsurl = {https://ui.adsabs.harvard.edu/abs/1990AJ....100.1091P},
	author = {{Peletier}, Reynier F. and {Davies}, Roger L. and {Illingworth}, Garth D. and {Davis}, Lindsey E. and {Cawson}, Michael},
	doi = {10.1086/115582},
	journal = {\aj},
	month = oct,
	pages = {1091},
	title = {{CCD Surface Photometry of Galaxies with Dynamical Data. II. UBR Photometry of 39 Elliptical Galaxies}},
	volume = {100},
	year = 1990}

@article{sandage72,
	adsurl = {https://ui.adsabs.harvard.edu/abs/1972ApJ...176...21S},
	author = {{Sandage}, Allan},
	doi = {10.1086/151606},
	journal = {\apj},
	month = aug,
	pages = {21},
	title = {{Absolute Magnitudes of E and so Galaxies in the Virgo and Coma Clusters as a Function of U - B Color}},
	volume = {176},
	year = 1972}

@article{suess19b,
	adsurl = {https://ui.adsabs.harvard.edu/abs/2019ApJ...877..103S},
	archiveprefix = {arXiv},
	author = {{Suess}, Katherine A. and {Kriek}, Mariska and {Price}, Sedona H. and {Barro}, Guillermo},
	doi = {10.3847/1538-4357/ab1bda},
	eid = {103},
	eprint = {1904.10992},
	journal = {\apj},
	month = jun,
	number = {2},
	pages = {103},
	primaryclass = {astro-ph.GA},
	title = {{Half-mass Radii for {\ensuremath{\sim}}7000 Galaxies at 1.0 {\ensuremath{\leq}} z {\ensuremath{\leq}} 2.5: Most of the Evolution in the Mass-Size Relation Is Due to Color Gradients}},
	volume = {877},
	year = 2019}

@inproceedings{perrin14,
	adsurl = {https://ui.adsabs.harvard.edu/abs/2014SPIE.9143E..3XP},
	author = {{Perrin}, Marshall D. and {Sivaramakrishnan}, Anand and {Lajoie}, Charles-Philippe and {Elliott}, Erin and {Pueyo}, Laurent and {Ravindranath}, Swara and {Albert}, Lo{\"\i}c.},
	booktitle = {Space Telescopes and Instrumentation 2014: Optical, Infrared, and Millimeter Wave},
	doi = {10.1117/12.2056689},
	editor = {{Oschmann}, Jacobus M., Jr. and {Clampin}, Mark and {Fazio}, Giovanni G. and {MacEwen}, Howard A.},
	eid = {91433X},
	month = aug,
	pages = {91433X},
	series = {Society of Photo-Optical Instrumentation Engineers (SPIE) Conference Series},
	title = {{Updated point spread function simulations for JWST with WebbPSF}},
	volume = {9143},
	year = 2014}

@article{valentino23,
	adsurl = {https://ui.adsabs.harvard.edu/abs/2023ApJ...947...20V},
	archiveprefix = {arXiv},
	author = {{Valentino}, Francesco and {Brammer}, Gabriel and {Gould}, Katriona M.~L. and {Kokorev}, Vasily and {Fujimoto}, Seiji and {Jespersen}, Christian Kragh and {Vijayan}, Aswin P. and {Weaver}, John R. and {Ito}, Kei and {Tanaka}, Masayuki and {Ilbert}, Olivier and {Magdis}, Georgios E. and {Whitaker}, Katherine E. and {Faisst}, Andreas L. and {Gallazzi}, Anna and {Gillman}, Steven and {Gim{\'e}nez-Arteaga}, Clara and {G{\'o}mez-Guijarro}, Carlos and {Kubo}, Mariko and {Heintz}, Kasper E. and {Hirschmann}, Michaela and {Oesch}, Pascal and {Onodera}, Masato and {Rizzo}, Francesca and {Lee}, Minju and {Strait}, Victoria and {Toft}, Sune},
	doi = {10.3847/1538-4357/acbefa},
	eid = {20},
	eprint = {2302.10936},
	journal = {\apj},
	month = apr,
	number = {1},
	pages = {20},
	primaryclass = {astro-ph.GA},
	title = {{An Atlas of Color-selected Quiescent Galaxies at z > 3 in Public JWST Fields}},
	volume = {947},
	year = 2023}

@misc{dunlop21,
	adsurl = {https://ui.adsabs.harvard.edu/abs/2021jwst.prop.1837D},
	author = {{Dunlop}, James S. and {Abraham}, Roberto G. and {Ashby}, Matthew L.~N. and {Bagley}, Micaela and {Best}, Philip N. and {Bongiorno}, Angela and {Bouwens}, Rychard and {Bowler}, Rebecca A.~A. and {Brammer}, Gabriel and {Bremer}, Malcolm and {Calabro'}, Antonello and {Carnall}, Adam and {Castellano}, Marco and {Cirasuolo}, Michele and {Conselice}, Christopher and {Cullen}, Fergus and {Dave}, Romeel and {Dayal}, Pratika and {Dekel}, Avishai and {Dickinson}, Mark and {Duncan}, Kenneth James and {Elbaz}, David and {Ellis}, Richard S. and {Ferguson}, Harry C. and {Ferrara}, Andrea and {Finkelstein}, Steven L. and {Fontana}, Adriano and {Furlanetto}, Steven and {Fynbo}, Johan P.~U. and {Gallerani}, Simona and {Gardner}, Jonathan P. and {Giavalisco}, Mauro and {Grazian}, Andrea and {Grogin}, Norman and {Harikane}, Yuichi and {Hopkins}, Philip F. and {Ilbert}, Olivier and {Illingworth}, Garth D. and {Juneau}, Stephanie and {Jung}, Intae and {Kartaltepe}, Jeyhan and {Kassin}, Susan and {Kauffmann}, Olivier Benjamin and {Khochfar}, Sadegh and {Kirkpatrick}, Allison and {Kocevski}, Dale D. and {Koekemoer}, Anton M. and {Labbe}, Ivo and {Laporte}, Nicolas and {Larson}, Rebecca L. and {Lucas}, Ray A. and {Magee}, Daniel K. and {Mason}, Charlotte and {McCracken}, Henry Joy and {McLeod}, Derek and {McLure}, Ross and {Merlin}, Emiliano and {Mesinger}, Andrei and {Milvang-Jensen}, Bo and {Newman}, Jeffrey Allen and {Oesch}, Pascal and {Ouchi}, Masami and {Pacifici}, Camilla and {Papovich}, Casey and {Peacock}, John and {Peeples}, Molly and {Pentericci}, Laura and {Perez-Gonzalez}, Pablo G. and {Pirzkal}, Norbert and {Pope}, Alexandra and {Pye}, John P. and {Reddy}, Naveen A. and {Robertson}, Brant and {Salvato}, Mara and {Santini}, Paola and {Schaerer}, Daniel and {Shapley}, Alice E. and {Simons}, Raymond and {Smit}, Renske and {Smith}, Britton D. and {Snyder}, Greg and {Somerville}, Rachel S. and {Stanway}, Elizabeth R. and {Stefanon}, Mauro and {Tasca}, Lidia and {Tikkanen}, Tuomo and {Tresse}, Laurence and {Trump}, Jonathan R. and {Whitaker}, Katherine E. and {Wilkins}, Stephen Matthew and {Wright}, Gillian and {Wyithe}, J. Stuart B. and {van Dokkum}, Pieter and {van der Werf}, Paul},
	howpublished = {JWST Proposal. Cycle 1, ID. \#1837},
	month = mar,
	pages = {1837},
	title = {{PRIMER: Public Release IMaging for Extragalactic Research}},
	year = 2021}

@article{casey23,
	adsurl = {https://ui.adsabs.harvard.edu/abs/2023ApJ...954...31C},
	archiveprefix = {arXiv},
	author = {{Casey}, Caitlin M. and {Kartaltepe}, Jeyhan S. and {Drakos}, Nicole E. and {Franco}, Maximilien and {Harish}, Santosh and {Paquereau}, Louise and {Ilbert}, Olivier and {Rose}, Caitlin and {Cox}, Isabella G. and {Nightingale}, James W. and {Robertson}, Brant E. and {Silverman}, John D. and {Koekemoer}, Anton M. and {Massey}, Richard and {McCracken}, Henry Joy and {Rhodes}, Jason and {Akins}, Hollis B. and {Allen}, Natalie and {Amvrosiadis}, Aristeidis and {Arango-Toro}, Rafael C. and {Bagley}, Micaela B. and {Bongiorno}, Angela and {Capak}, Peter L. and {Champagne}, Jaclyn B. and {Chartab}, Nima and {Ch{\'a}vez Ortiz}, {\'O}scar A. and {Chworowsky}, Katherine and {Cooke}, Kevin C. and {Cooper}, Olivia R. and {Darvish}, Behnam and {Ding}, Xuheng and {Faisst}, Andreas L. and {Finkelstein}, Steven L. and {Fujimoto}, Seiji and {Gentile}, Fabrizio and {Gillman}, Steven and {Gould}, Katriona M.~L. and {Gozaliasl}, Ghassem and {Hayward}, Christopher C. and {He}, Qiuhan and {Hemmati}, Shoubaneh and {Hirschmann}, Michaela and {Jahnke}, Knud and {Jin}, Shuowen and {Khostovan}, Ali Ahmad and {Kokorev}, Vasily and {Lambrides}, Erini and {Laigle}, Clotilde and {Larson}, Rebecca L. and {Leung}, Gene C.~K. and {Liu}, Daizhong and {Liaudat}, Tobias and {Long}, Arianna S. and {Magdis}, Georgios and {Mahler}, Guillaume and {Mainieri}, Vincenzo and {Manning}, Sinclaire M. and {Maraston}, Claudia and {Martin}, Crystal L. and {McCleary}, Jacqueline E. and {McKinney}, Jed and {McPartland}, Conor J.~R. and {Mobasher}, Bahram and {Pattnaik}, Rohan and {Renzini}, Alvio and {Rich}, R. Michael and {Sanders}, David B. and {Sattari}, Zahra and {Scognamiglio}, Diana and {Scoville}, Nick and {Sheth}, Kartik and {Shuntov}, Marko and {Sparre}, Martin and {Suzuki}, Tomoko L. and {Talia}, Margherita and {Toft}, Sune and {Trakhtenbrot}, Benny and {Urry}, C. Megan and {Valentino}, Francesco and {Vanderhoof}, Brittany N. and {Vardoulaki}, Eleni and {Weaver}, John R. and {Whitaker}, Katherine E. and {Wilkins}, Stephen M. and {Yang}, Lilan and {Zavala}, Jorge A.},
	doi = {10.3847/1538-4357/acc2bc},
	eid = {31},
	eprint = {2211.07865},
	journal = {\apj},
	month = sep,
	number = {1},
	pages = {31},
	primaryclass = {astro-ph.GA},
	title = {{COSMOS-Web: An Overview of the JWST Cosmic Origins Survey}},
	volume = {954},
	year = 2023}

@manual{ng22,
	author = {{Ng}, Pin and {Maechler}, Martin},
	note = {R package version 1.3-5.},
	title = {COBS -- Constrained B-splines (Sparse matrix based)},
	url = {https://CRAN.R-project.org/package=cobs},
	year = {2022}}

@article{ng07,
	author = {{Ng}, Pin and {Maechler}, Martin},
	doi = {10.1177/1471082X0700700403},
	journal = {Statistical Modelling},
	number = {4},
	pages = {315--328},
	title = {A Fast and Efficient Implementation of Qualitatively Constrained Quantile Smoothing Splines},
	volume = {7},
	year = {2007}}

@article{leja22,
	adsurl = {https://ui.adsabs.harvard.edu/abs/2022ApJ...936..165L},
	archiveprefix = {arXiv},
	author = {{Leja}, Joel and {Speagle}, Joshua S. and {Ting}, Yuan-Sen and {Johnson}, Benjamin D. and {Conroy}, Charlie and {Whitaker}, Katherine E. and {Nelson}, Erica J. and {van Dokkum}, Pieter and {Franx}, Marijn},
	doi = {10.3847/1538-4357/ac887d},
	eid = {165},
	eprint = {2110.04314},
	journal = {\apj},
	month = sep,
	number = {2},
	pages = {165},
	primaryclass = {astro-ph.GA},
	title = {{A New Census of the 0.2 < z < 3.0 Universe. II. The Star-forming Sequence}},
	volume = {936},
	year = 2022}

@article{Finkelstein23,
	adsurl = {https://ui.adsabs.harvard.edu/abs/2023ApJ...946L..13F},
	archiveprefix = {arXiv},
	author = {{Finkelstein}, Steven L. and {Bagley}, Micaela B. and {Ferguson}, Henry C. and {Wilkins}, Stephen M. and {Kartaltepe}, Jeyhan S. and {Papovich}, Casey and {Yung}, L.~Y. Aaron and {Arrabal Haro}, Pablo and {Behroozi}, Peter and {Dickinson}, Mark and {Kocevski}, Dale D. and {Koekemoer}, Anton M. and {Larson}, Rebecca L. and {Le Bail}, Aur{\'e}lien and {Morales}, Alexa M. and {P{\'e}rez-Gonz{\'a}lez}, Pablo G. and {Burgarella}, Denis and {Dav{\'e}}, Romeel and {Hirschmann}, Michaela and {Somerville}, Rachel S. and {Wuyts}, Stijn and {Bromm}, Volker and {Casey}, Caitlin M. and {Fontana}, Adriano and {Fujimoto}, Seiji and {Gardner}, Jonathan P. and {Giavalisco}, Mauro and {Grazian}, Andrea and {Grogin}, Norman A. and {Hathi}, Nimish P. and {Hutchison}, Taylor A. and {Jha}, Saurabh W. and {Jogee}, Shardha and {Kewley}, Lisa J. and {Kirkpatrick}, Allison and {Long}, Arianna S. and {Lotz}, Jennifer M. and {Pentericci}, Laura and {Pierel}, Justin D.~R. and {Pirzkal}, Nor and {Ravindranath}, Swara and {Ryan}, Russell E. and {Trump}, Jonathan R. and {Yang}, Guang and {Bhatawdekar}, Rachana and {Bisigello}, Laura and {Buat}, V{\'e}ronique and {Calabr{\`o}}, Antonello and {Castellano}, Marco and {Cleri}, Nikko J. and {Cooper}, M.~C. and {Croton}, Darren and {Daddi}, Emanuele and {Dekel}, Avishai and {Elbaz}, David and {Franco}, Maximilien and {Gawiser}, Eric and {Holwerda}, Benne W. and {Huertas-Company}, Marc and {Jaskot}, Anne E. and {Leung}, Gene C.~K. and {Lucas}, Ray A. and {Mobasher}, Bahram and {Pandya}, Viraj and {Tacchella}, Sandro and {Weiner}, Benjamin J. and {Zavala}, Jorge A.},
	doi = {10.3847/2041-8213/acade4},
	eid = {L13},
	eprint = {2211.05792},
	journal = {\apjl},
	month = mar,
	number = {1},
	pages = {L13},
	primaryclass = {astro-ph.GA},
	title = {{CEERS Key Paper. I. An Early Look into the First 500 Myr of Galaxy Formation with JWST}},
	volume = {946},
	year = 2023}

@article{miller23,
	adsurl = {https://ui.adsabs.harvard.edu/abs/2023ApJ...945..155M},
	archiveprefix = {arXiv},
	author = {{Miller}, Tim B. and {van Dokkum}, Pieter and {Mowla}, Lamiya},
	doi = {10.3847/1538-4357/acbc74},
	eid = {155},
	eprint = {2207.05895},
	journal = {\apj},
	month = mar,
	number = {2},
	pages = {155},
	primaryclass = {astro-ph.GA},
	title = {{Color Gradients and Half-mass Radii of Galaxies Out to z = 2 in the CANDELS/3D-HST Fields: Further Evidence for Important Differences in the Evolution of Mass-weighted and Light-weighted Sizes}},
	volume = {945},
	year = 2023}

@article{tadaki17,
	adsurl = {https://ui.adsabs.harvard.edu/abs/2017ApJ...834..135T},
	archiveprefix = {arXiv},
	author = {{Tadaki}, Ken-ichi and {Genzel}, Reinhard and {Kodama}, Tadayuki and {Wuyts}, Stijn and {Wisnioski}, Emily and {F{\"o}rster Schreiber}, Natascha M. and {Burkert}, Andreas and {Lang}, Philipp and {Tacconi}, Linda J. and {Lutz}, Dieter and {Belli}, Sirio and {Davies}, Richard I. and {Hatsukade}, Bunyo and {Hayashi}, Masao and {Herrera-Camus}, Rodrigo and {Ikarashi}, Soh and {Inoue}, Shigeki and {Kohno}, Kotaro and {Koyama}, Yusei and {Mendel}, J. Trevor and {Nakanishi}, Kouichiro and {Shimakawa}, Rhythm and {Suzuki}, Tomoko L. and {Tamura}, Yoichi and {Tanaka}, Ichi and {{\"U}bler}, Hannah and {Wilman}, Dave J.},
	doi = {10.3847/1538-4357/834/2/135},
	eid = {135},
	eprint = {1608.05412},
	journal = {\apj},
	month = jan,
	number = {2},
	pages = {135},
	primaryclass = {astro-ph.GA},
	title = {{Bulge-forming Galaxies with an Extended Rotating Disk at z \raisebox{-0.5ex}\textasciitilde 2}},
	volume = {834},
	year = 2017}

@article{suess22,
	adsurl = {https://ui.adsabs.harvard.edu/abs/2022ApJ...937L..33S},
	archiveprefix = {arXiv},
	author = {{Suess}, Katherine A. and {Bezanson}, Rachel and {Nelson}, Erica J. and {Setton}, David J. and {Price}, Sedona H. and {van Dokkum}, Pieter and {Brammer}, Gabriel and {Labb{\'e}}, Ivo and {Leja}, Joel and {Miller}, Tim B. and {Robertson}, Brant and {Wel}, Arjen van der and {Weaver}, John R. and {Whitaker}, Katherine E.},
	doi = {10.3847/2041-8213/ac8e06},
	eid = {L33},
	eprint = {2207.10655},
	journal = {\apjl},
	month = oct,
	number = {2},
	pages = {L33},
	primaryclass = {astro-ph.GA},
	title = {{Rest-frame Near-infrared Sizes of Galaxies at Cosmic Noon: Objects in JWST's Mirror Are Smaller than They Appeared}},
	volume = {937},
	year = 2022}

@article{de-jong96a,
	adsurl = {https://ui.adsabs.harvard.edu/abs/1996A&A...313...45D},
	archiveprefix = {arXiv},
	author = {{de Jong}, R.~S.},
	doi = {10.48550/arXiv.astro-ph/9601005},
	eprint = {astro-ph/9601005},
	journal = {\aap},
	month = sep,
	pages = {45-64},
	primaryclass = {astro-ph},
	title = {{Near-infrared and optical broadband surface photometry of 86 face-on disk dominated galaxies. III. The statistics of the disk and bulge parameters.}},
	volume = {313},
	year = 1996}

@article{barbary16,
	author = {Kyle Barbary},
	journal = {Journal of Open Source Software},
	number = {6},
	pages = {58},
	title = {SEP: Source Extractor as a library},
	volume = {1},
	year = {2016}}

@article{tacchella16,
	adsurl = {https://ui.adsabs.harvard.edu/abs/2016MNRAS.458..242T},
	archiveprefix = {arXiv},
	author = {{Tacchella}, Sandro and {Dekel}, Avishai and {Carollo}, C. Marcella and {Ceverino}, Daniel and {DeGraf}, Colin and {Lapiner}, Sharon and {Mandelker}, Nir and {Primack}, Joel R.},
	doi = {10.1093/mnras/stw303},
	eprint = {1509.00017},
	journal = {\mnras},
	month = may,
	number = {1},
	pages = {242-263},
	primaryclass = {astro-ph.GA},
	title = {{Evolution of density profiles in high-z galaxies: compaction and quenching inside-out}},
	volume = {458},
	year = 2016}

@article{tadaki20,
	adsurl = {https://ui.adsabs.harvard.edu/abs/2020ApJ...901...74T},
	archiveprefix = {arXiv},
	author = {{Tadaki}, Ken-ichi and {Belli}, Sirio and {Burkert}, Andreas and {Dekel}, Avishai and {F{\"o}rster Schreiber}, Natascha M. and {Genzel}, Reinhard and {Hayashi}, Masao and {Herrera-Camus}, Rodrigo and {Kodama}, Tadayuki and {Kohno}, Kotaro and {Koyama}, Yusei and {Lee}, Minju M. and {Lutz}, Dieter and {Mowla}, Lamiya and {Nelson}, Erica J. and {Renzini}, Alvio and {Suzuki}, Tomoko L. and {Tacconi}, Linda J. and {{\"U}bler}, Hannah and {Wisnioski}, Emily and {Wuyts}, Stijn},
	doi = {10.3847/1538-4357/abaf4a},
	eid = {74},
	eprint = {2009.01976},
	journal = {\apj},
	month = sep,
	number = {1},
	pages = {74},
	primaryclass = {astro-ph.GA},
	title = {{Structural Evolution in Massive Galaxies at z {\ensuremath{\sim}} 2}},
	volume = {901},
	year = 2020}

@article{nelson16,
	adsurl = {https://ui.adsabs.harvard.edu/abs/2016ApJ...817L...9N},
	archiveprefix = {arXiv},
	author = {{Nelson}, Erica June and {van Dokkum}, Pieter G. and {Momcheva}, Ivelina G. and {Brammer}, Gabriel B. and {Wuyts}, Stijn and {Franx}, Marijn and {F{\"o}rster Schreiber}, Natascha M. and {Whitaker}, Katherine E. and {Skelton}, Rosalind E.},
	doi = {10.3847/2041-8205/817/1/L9},
	eid = {L9},
	eprint = {1511.04443},
	journal = {\apjl},
	month = jan,
	number = {1},
	pages = {L9},
	primaryclass = {astro-ph.GA},
	title = {{Spatially Resolved Dust Maps from Balmer Decrements in Galaxies at z \raisebox{-0.5ex}\textasciitilde 1.4}},
	volume = {817},
	year = 2016}

@article{tacchella18,
	adsurl = {https://ui.adsabs.harvard.edu/abs/2018ApJ...859...56T},
	archiveprefix = {arXiv},
	author = {{Tacchella}, S. and {Carollo}, C.~M. and {F{\"o}rster Schreiber}, N.~M. and {Renzini}, A. and {Dekel}, A. and {Genzel}, R. and {Lang}, P. and {Lilly}, S.~J. and {Mancini}, C. and {Onodera}, M. and {Tacconi}, L.~J. and {Wuyts}, S. and {Zamorani}, G.},
	doi = {10.3847/1538-4357/aabf8b},
	eid = {56},
	eprint = {1704.00733},
	journal = {\apj},
	month = may,
	number = {1},
	pages = {56},
	primaryclass = {astro-ph.GA},
	title = {{Dust Attenuation, Bulge Formation, and Inside-out Quenching of Star Formation in Star-forming Main Sequence Galaxies at z {\ensuremath{\sim}} 2}},
	volume = {859},
	year = 2018}

@misc{Finkelstein17,
	adsurl = {https://ui.adsabs.harvard.edu/abs/2017jwst.prop.1345F},
	author = {{Finkelstein}, Steven L. and {Dickinson}, Mark and {Ferguson}, Harry C. and {Grazian}, Andrea and {Grogin}, Norman and {Kartaltepe}, Jeyhan and {Kewley}, Lisa and {Kocevski}, Dale D. and {Koekemoer}, Anton M. and {Lotz}, Jennifer and {Papovich}, Casey and {Pentericci}, Laura and {Perez-Gonzalez}, Pablo G. and {Pirzkal}, Norbert and {Ravindranath}, Swara and {Somerville}, Rachel S. and {Trump}, Jonathan R. and {Wilkins}, Stephen Matthew},
	howpublished = {JWST Proposal ID 1345. Cycle 0 Early Release Science},
	month = nov,
	pages = {1345},
	title = {{The Cosmic Evolution Early Release Science (CEERS) Survey}},
	year = 2017}

@article{barisic20,
	adsurl = {https://ui.adsabs.harvard.edu/abs/2020ApJ...903..146B},
	archiveprefix = {arXiv},
	author = {{Bari{\v{s}}i{\'c}}, Ivana and {Pacifici}, Camilla and {van der Wel}, Arjen and {Straatman}, Caroline and {Bell}, Eric F. and {Bezanson}, Rachel and {Brammer}, Gabriel and {D'Eugenio}, Francesco and {Franx}, Marijn and {van Houdt}, Josha and {Maseda}, Michael V. and {Muzzin}, Adam and {Sobral}, David and {Wu}, Po-Feng},
	doi = {10.3847/1538-4357/abba37},
	eid = {146},
	eprint = {2010.01147},
	journal = {\apj},
	month = nov,
	number = {2},
	pages = {146},
	primaryclass = {astro-ph.GA},
	title = {{Dust Attenuation Curves at z {\ensuremath{\sim}} 0.8 from LEGA-C: Precise Constraints on the Slope and 2175{\v{S}}Bump Strength}},
	volume = {903},
	year = 2020}

@article{conroy09,
	adsurl = {https://ui.adsabs.harvard.edu/abs/2009ApJ...699..486C},
	archiveprefix = {arXiv},
	author = {{Conroy}, Charlie and {Gunn}, James E. and {White}, Martin},
	doi = {10.1088/0004-637X/699/1/486},
	eprint = {0809.4261},
	journal = {\apj},
	month = jul,
	number = {1},
	pages = {486-506},
	primaryclass = {astro-ph},
	title = {{The Propagation of Uncertainties in Stellar Population Synthesis Modeling. I. The Relevance of Uncertain Aspects of Stellar Evolution and the Initial Mass Function to the Derived Physical Properties of Galaxies}},
	volume = {699},
	year = 2009}

@article{haussler13,
	adsurl = {https://ui.adsabs.harvard.edu/abs/2013MNRAS.430..330H},
	archiveprefix = {arXiv},
	author = {{H{\"a}u{\ss}ler}, Boris and {Bamford}, Steven P. and {Vika}, Marina and {Rojas}, Alex L. and {Barden}, Marco and {Kelvin}, Lee S. and {Alpaslan}, Mehmet and {Robotham}, Aaron S.~G. and {Driver}, Simon P. and {Baldry}, I.~K. and {Brough}, Sarah and {Hopkins}, Andrew M. and {Liske}, Jochen and {Nichol}, Robert C. and {Popescu}, Cristina C. and {Tuffs}, Richard J.},
	doi = {10.1093/mnras/sts633},
	eprint = {1212.3332},
	journal = {\mnras},
	month = mar,
	number = {1},
	pages = {330-369},
	primaryclass = {astro-ph.CO},
	title = {{MegaMorph - multiwavelength measurement of galaxy structure: complete S{\'e}rsic profile information from modern surveys}},
	volume = {430},
	year = 2013}

@article{zhang19,
	adsurl = {https://ui.adsabs.harvard.edu/abs/2019MNRAS.484.5170Z},
	archiveprefix = {arXiv},
	author = {{Zhang}, Haowen and {Primack}, Joel R. and {Faber}, S.~M. and {Koo}, David C. and {Dekel}, Avishai and {Chen}, Zhu and {Ceverino}, Daniel and {Chang}, Yu-Yen and {Fang}, Jerome J. and {Guo}, Yicheng and {Lin}, Lin and {Wel}, Arjen van der},
	doi = {10.1093/mnras/stz339},
	eprint = {1805.12331},
	journal = {\mnras},
	month = apr,
	number = {4},
	pages = {5170-5191},
	primaryclass = {astro-ph.GA},
	title = {{The evolution of galaxy shapes in CANDELS: from prolate to discy}},
	volume = {484},
	year = 2019}

@article{bertin96,
	adsurl = {http://adsabs.harvard.edu/abs/1996A%26AS..117..393B},
	author = {{Bertin}, E. and {Arnouts}, S.},
	doi = {10.1051/aas:1996164},
	journal = {\aaps},
	month = jun,
	pages = {393-404},
	title = {{SExtractor: Software for source extraction.}},
	volume = 117,
	year = 1996}

\begin{appendix}

\section{Effect of residual flux on the measured color gradients}\label{appendix:residuals}
    \begin{figure*}[ht!]
            \centering
        \includegraphics[scale=0.5]{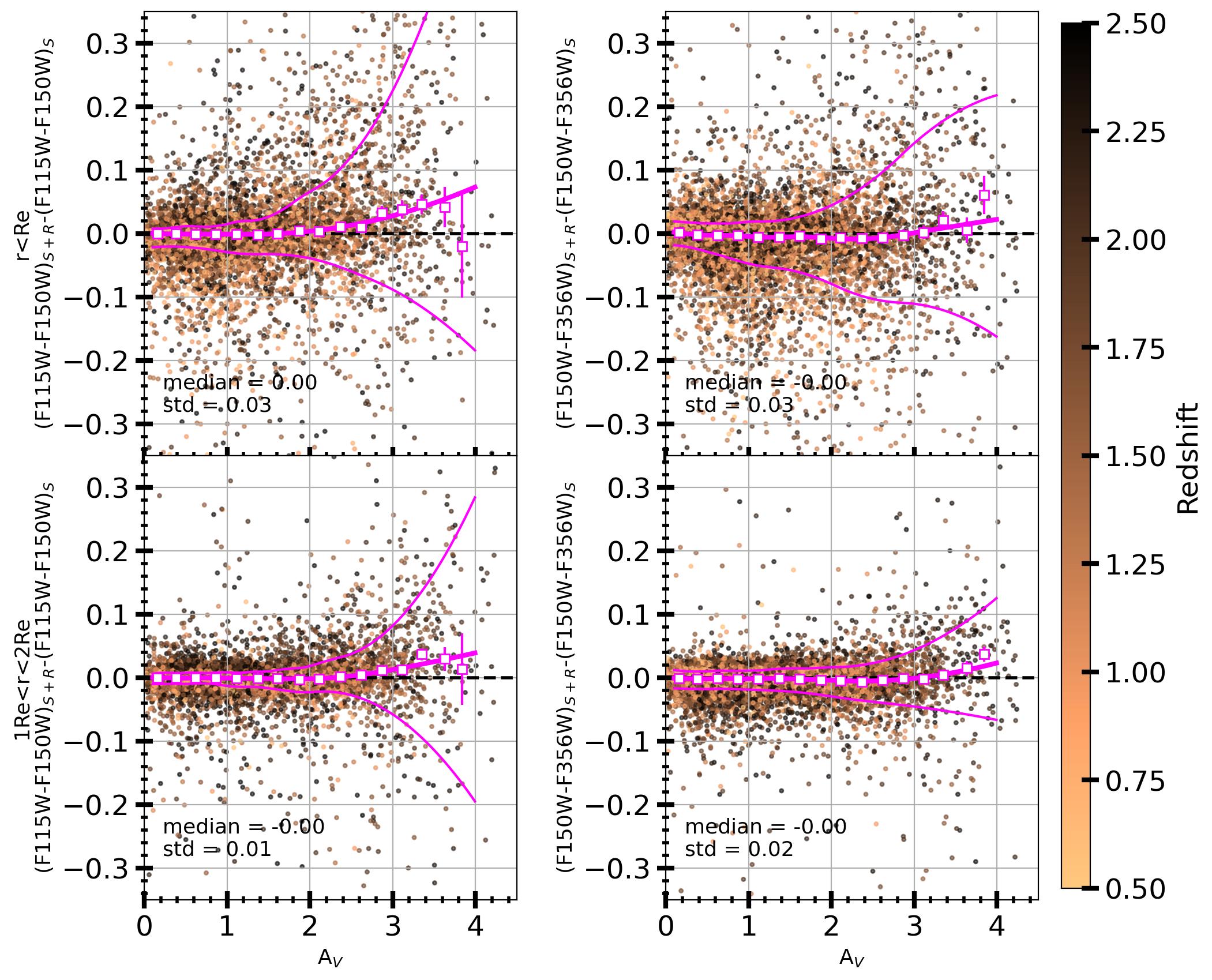}
            \caption{Difference between the flux difference (i.e. color gradient) between F115W$-$F150W (first column) and F150W$-$F356W (second column) computed using photometry from the S\'ersic profile adding back the fit residuals (S+R) or just from the S\'ersic profiles (S) measured within the effective radius $R_{\text{e}, F444W}$ (first row) or between $1-2~R_{\text{e}, F444W}$ (second row). These are presented as a function of the global $A_V$ and color-coded with the galaxy redshift. White squares with magenta contour show the median difference in $A_V$ bins of width $0.2~\text{mag}$, while error bars show the statistical uncertainty ($\sigma/\sqrt{N}$). Solid magenta lines show the 16-50-84 percentiles computed with the \textsc{cobs} library \citep{ng07,ng22}, which allows for a smoothed combination of a spline regression and quantile regression. In the bottom left corner of each panel we report the median color difference and the standard deviation.
            }
        \label{fig:appendix_ColorComparison}
    \end{figure*}

    The color gradients investigated in this paper are computed based on the single S\'ersic profiles from GalfitM, without a residual correction as first employed by \cite{szomoru10}. In this appendix, we verify the impact of the residuals on the retrieved colors.
    We measured the residual fluxes within $R_{\text{e}, F444W}$ and between $1-2 R_{\text{e}, F444W}$ for three filters (F115W, F150W and F356W), which approximately cover the $U$, $V$, and $J$ bands at $z=2$. We then compute F115W$-$F150W and F150W$-$F356W colors from the S\'ersic profiles (S) and from the S\'ersic profiles plus the fit residuals (S+R). The difference between S+R color and S color is shown in Figure \ref{fig:appendix_ColorComparison} as a function of $A_V$ and color-coded with the redshift.
    The median difference between colors measured within $R_{\text{e}}$ and between $1-2~R_{\text{e}, F444W}$ is negligible, with a maximum standard deviation of $0.03~\text{mag}$. The maximum standard deviation occurs for colors measured in the core of galaxies, where residual corrections are sensitive to errors in the PSF model. 
    The 16-84 percentile range is increased for highly attenuated galaxies, for which the gradient estimates are also the most uncertain due to the small flux in F115W: the effect of the residuals is comparable with the uncertainties.
    Concerning the outer annuli (bottom row of the figure), we see that the residuals play no significant role, with only a mild increase in scatter for high $A_V$ in the color F115W$-$F150W.
    We find no significant dependence on redshift or stellar mass.
    
    We note that for individual images the residuals are larger than for the colors: the covariance between the residuals in different filters reduces the impact on our color gradient estimates. Since residual corrections are very small (Fig.~\ref{fig:appendix_ColorComparison}) and can introduce new errors (PSF uncertainties affect the centers; noise peaks affect the low surface brightness outer parts), we choose to omit the residual corrections in this paper.

\section{Estimate of the color gradient's uncertainties}\label{appendix:errors}

    \begin{figure}[ht!]
            \centering
        \includegraphics[scale=0.35]{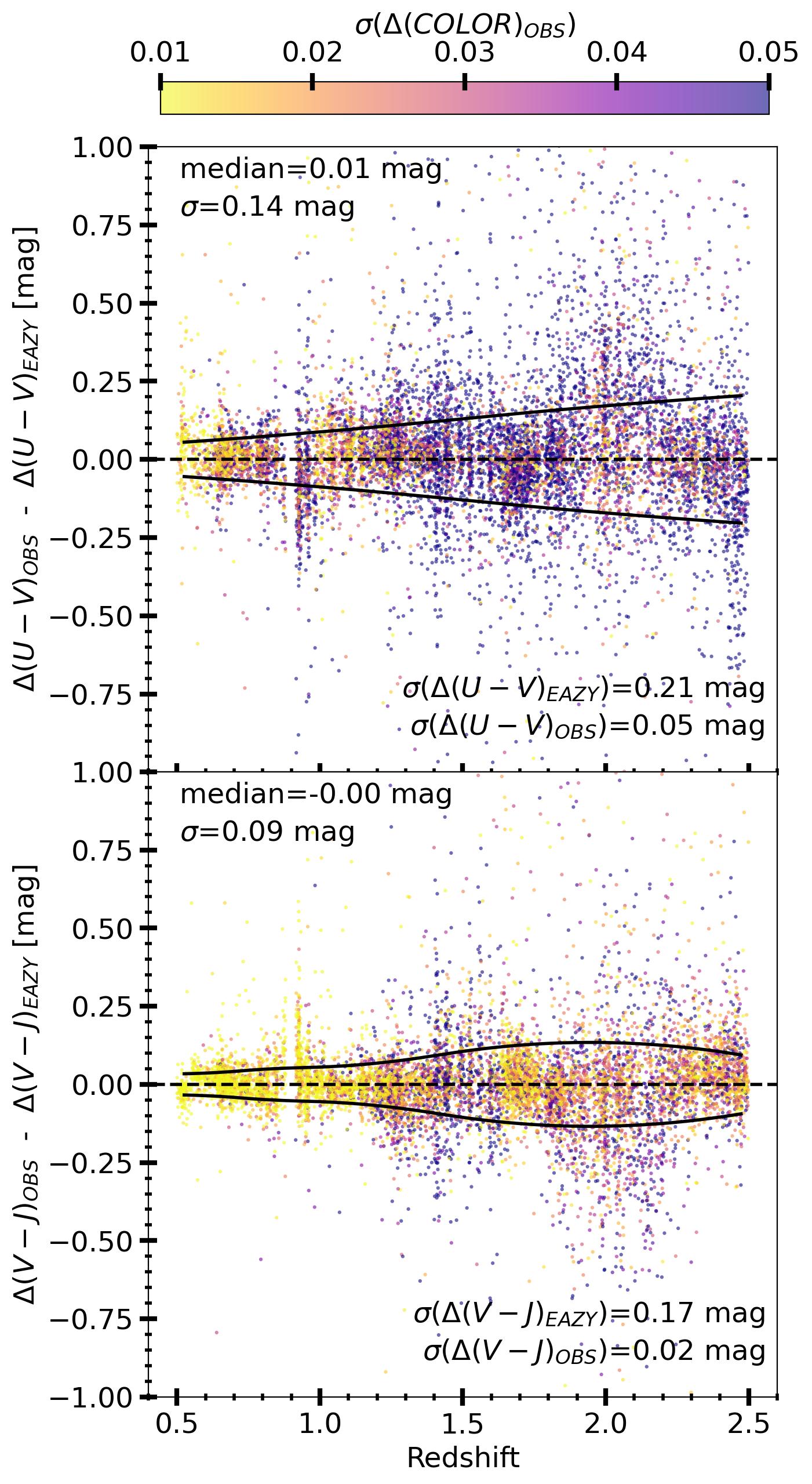}
            \caption{Comparison between the color gradients retrieved with \textsc{Eazy} rest-frame fluxes and observed fluxes as a function of redshift and color coded with the uncertainty on the color gradient retrieved using observed fluxes. The top panel shows the ${\textit{U}} - {\textit{V}}$ color difference, while the bottom panel shows the ${\textit{V}} - {\textit{J}}$ color difference. In the top left corner of each panel we present the median and standard deviation of the color gradient difference, while in the bottom right corner we show the median uncertainty on the two color gradients. Solid black lines show the standard deviation of the difference between the two color gradients as a function of redshift.
            }
        \label{fig:ColorComparison}
    \end{figure}
                    
    Color gradients presented in this work are computed by retrieving rest-frame \textit{U}, \textit{V}, \textit{J} fluxes via SED fitting with the code \textsc{Eazy}.
    Random uncertainties are given by a combination of the observed flux uncertainties and model uncertainties - the underlying SED. Uncertainties provided by \textsc{Eazy} on the rest-frame fluxes include both. When computing color gradients, the covariant terms of the uncertainty on each band's flux cancel out with the net effect that the final uncertainty is expected to be smaller than the simple quadrature sum of the uncertainties of all fluxes involved in the color gradient. In fact, covariant terms originate from the SED model, which affects the same way rest-frame fluxes computed at a fixed aperture.
    On the other hand, uncertainties on color gradients computed from observed fluxes are just given by the quadrature sum of the uncertainties on all the fluxes involved since no convolution with an SED model is involved in their determination. 
    
    To estimate the true uncertainty on color gradients, we compared the color gradients derived using rest-frame colors from \textsc{Eazy} (the same used throughout the paper) with color gradients computed using observed fluxes in the filters closest to the rest-frame bands investigated. 
    The total scatter of the former ($\sigma^2_{Eazy}$) can be written as 
    \begin{equation}
        \sigma^2_{Eazy} = \sigma^2_{flux} + \sigma^2_{covariant}
    \end{equation}
    The constraint set in Section 2.3 that observations must overlap at least 50\% of the rest-frame bands, combined with the statements of \cite{brammer11} who found colors retrieved with \textsc{Eazy} to accurately match colors retrieved by interpolating fluxes in nearby observed bands (when these overlap sufficiently with the rest-frame band), makes any bandpass mismatch errors negligible. The total scatter of the second ($\sigma^2_{OBS}$) can be written as 
    \begin{equation}
        \sigma^2_{OBS} = \sigma^2_{flux} + \sigma^2_{\lambda}
    \end{equation}
    with $\sigma^2_{flux}$ the scatter due to flux uncertainties, $\sigma^2_{covariant}$ the scatter due to covariance terms and $\sigma^2_{\lambda}$ the scatter due to the difference in band coverage between the rest-frame band investigated and the closest filter.
    Since the sample is selected such that at least $50\%$ of the area of the rest-frame band overlaps with observations, the scatter due to this term can be considered negligible compared to the others.
    
    The scatter in the comparison between the two color gradients is then representative of the covariant term. Hence, the real uncertainty on \textsc{Eazy} color gradients used in this work can be computed as $\sigma_{ColorGrad}=\sqrt{\sigma^2_{OBS}+\sigma^2_{covariant}(z)}$ where the latter term is represented by the scatter in Figure \ref{fig:ColorComparison} and is expressed a function of redshift.

    To account also for the uncertainties of $R_\text{e}$ on the color gradients, we add in quadrature to $\sigma_{ColorGrad}$ a new term $\sigma_{Sersic}$ computed as the ratio between the fraction of light enclosed in the S\'ersic profile tracing the \textit{U} band up to $R_{e,V}$ that is the effective radius measured in the \textit{V} band. The same is done for the \textit{V} and \textit{J} bands.
    This term accounts for the uncertainty induced by retrieving photometry from several different modeled S\'ersic profiles and has a median value of $\sigma_{Sersic,U-V}\approx0.07~\text{mag}$ and $\sigma_{Sersic,V-J}\approx0.09~\text{mag}$, comparable or smaller than the scatter shown in Figure \ref{fig:ColorComparison}.

\section{Color gradients and the wavelength dependence of size} \label{sec:sizes}
    
    \begin{figure*}[ht!]
            \centering
        \includegraphics[scale=0.38]
        {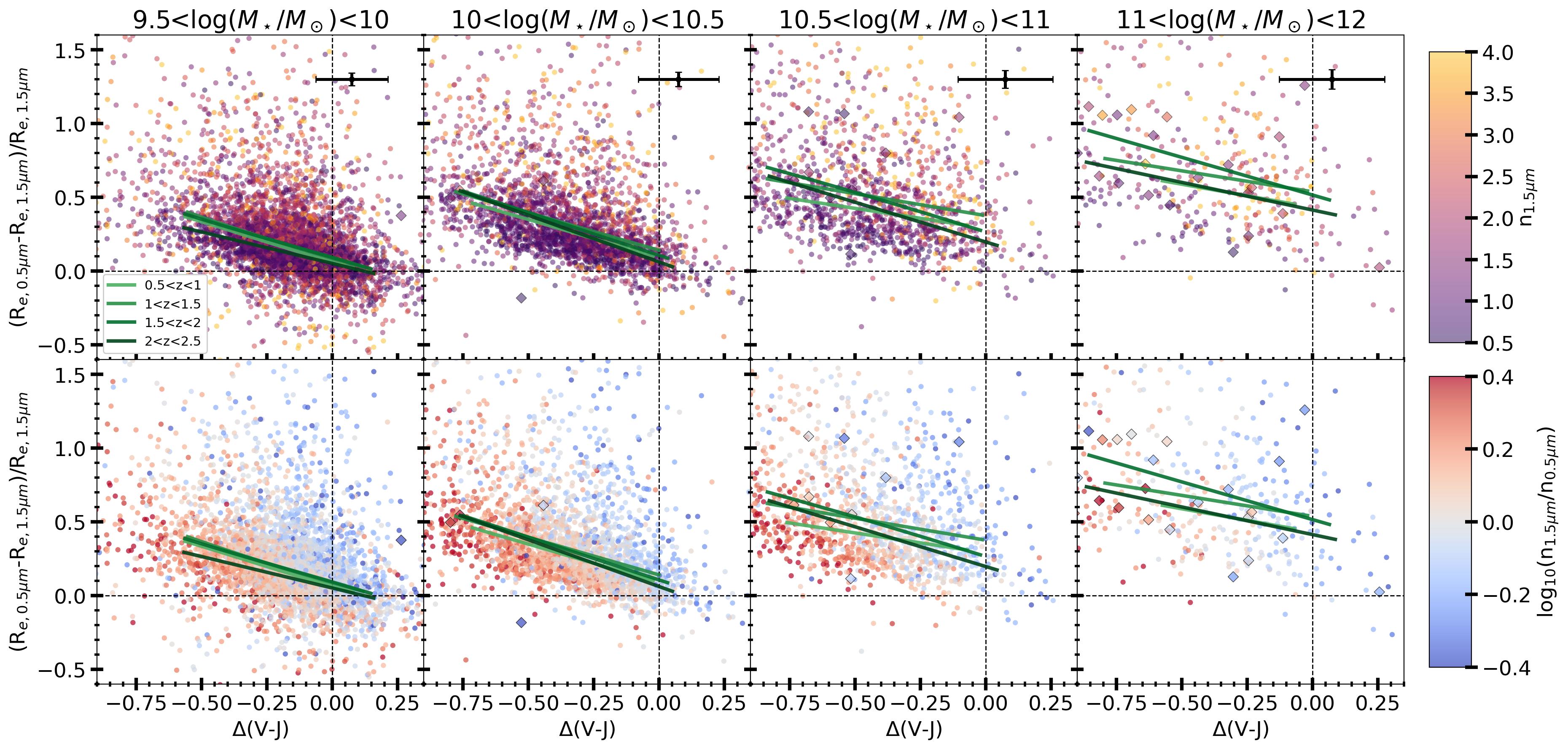}
            \caption{${\textit{V}} - {\textit{J}}$ color gradient vs relative size variation in four stellar mass bins. The two rows show the same data but color-coded respectively with the S\'ersic index at $1.5~\mu\text{m}$ (top row) and the logarithm difference of S\'ersic indices at rest-frame $1.5~\mu{\text{m}}$ and $0.5~\mu{\text{m}}$ (lower row). Just star-forming galaxies are shown. Diamonds show sub-mm-selected galaxies and are color-coded following the same color scheme of the other galaxies.
            Colored green lines show the running median in redshift bins. In the top right corner, we present the median uncertainties.
            Massive star-forming galaxies have systematically stronger size variation with wavelength and stronger color gradients than low-mass galaxies. The variation of the S\'ersic index with wavelength correlates more strongly with the color gradient than with the size gradient.}
        \label{fig:Size-ColGrad}
    \end{figure*}
    
A color gradient and a size difference measured at different wavelengths are (nearly) the same thing, especially when both derive from the same light profile fits. In fact, such size differences are often referred to as color gradients \citep[i.e., ][]{suess22,cutler24,ito24,ormerod24,martorano24,clausen25}. Specifically, in our previous work \cite{martorano24}, we found a median optical-to-near-IR size ratio of $0.14~{\text{dex}}$, and up to $0.25~{\text{dex}}$ for massive star-forming galaxies at $z>1.5$. These results precisely mirror the color gradient trends analyzed in this paper, which aims at interpreting the origin of the color gradients (attenuation) rather than its effect on the measured structural evolution of galaxies.
It is still useful to compare the color gradients, in this paper defined as the color difference between the outskirts (1 and 2$R_\text{e}$) and the center (inside $R_\text{e}$), with the size difference between the optical (rest-frame $0.5~\mu\text{m}$) and near-IR (rest-frame $1.5~\mu\text{m}$) from \cite{martorano24}, even if both derive from the same S\'ersic profile fits. In Figure \ref{fig:Size-ColGrad} we compare, for star-forming galaxies, the size differences with the ${\textit{V}} - {\textit{J}}$ color gradients.

There exists a strong correlation between the stellar mass and the size variation: low-mass galaxies have a median size variation of $\approx25\%$ while massive galaxies have median variations of $\approx65\%$, as presented in \cite{martorano24}. No clear redshift trend is recovered for any of the stellar mass bin investigated.
By construction, this size difference closely correlates with our $\Delta(V-J)$, but there is substantial scatter that originates from variations in S\'ersic index, that is the distribution of light within the two large apertures used to define the color gradient. A larger S\'ersic index produces a larger size difference at a fixed color gradient (top row in Figure \ref{fig:Size-ColGrad}). In addition, if the S\'ersic index varies with wavelength \citep[i.e.,][]{kelvin12, vika13, kennedy15, martorano23, martorano25}, then galaxies with strong color gradients may show just a mild difference in size and, vice versa, galaxies with mild color gradients may show strong size gradients (lower row in Figure \ref{fig:Size-ColGrad}). 
The take-away message is that radial variations in color do not necessarily lead to net color gradient expressed as a summary statistic (here, comparing the average color between 1 and 2 $R_\text{e}$ with that inside $R_\text{e}$ as measured in F444W). Conversely, the lack of a color gradient as defined here, does not necessarily imply that the color is the same at all radii.
The lack of a clear redshift trend in this figure proves that the size variation as a function of redshift presented in \cite{martorano24} is comparable to that presented in this work.

Sub-mm galaxies in our sample exhibit both strong color gradients and size variations (confirming results presented in \citealt{ren25}) that closely align with the median trends. These galaxies have median $n_{1.5\mu{\text{m}}}\sim1.3$ and $n_{0.5\mu{\text{m}}}\sim1.1$ confirming findings of \cite{price25, gillman24} and several others. As in \cite{price25}, we find more concentrated sub-mm galaxies to have stronger size variation, suggesting galaxies with the most compact light distributions also have the most concentrated dust.
\end{appendix}

\end{document}